\documentclass[aps,pra,twocolumn,groupedaddress,floatfix,showpacs,longbibliography]{revtex4-2}
\usepackage{graphicx}
\usepackage{amsmath}
\usepackage{color}

\begin{document}

\title{Revealing the Spin Hydrodynamics of a Spin-Imbalanced Unitary Fermi Gas}

\author{DeChao Zhang$^{1}$, Johannes Lang$^{2}$, and J. E. Thomas$^{1\ast}$}

\affiliation{$^{1}$Department of  Physics, North Carolina State University, Raleigh, NC, USA\\
$^{2}$Institut f\"ur Theoretische Physik, Universit\"at zu K\"oln, Cologne, Germany}

\date{\today}

\begin{abstract}
Hydrodynamics governs diverse collective phenomena in nature, from the expansion of a quark-gluon plasma to viscous electron flow in quantum materials.
 In strongly interacting hydrodynamic fluids,  the  transport of spin remains poorly understood. Here, we investigate spin transport in the hydrodynamic regime of a spin-imbalanced unitary Fermi gas confined in a uniform optical box. By quenching a spatially periodic optical potential that modulates both the density and spin polarization,  we observe the relaxation of the many-body system, which determines both the spin diffusivity and the spin Seebeck/Peltier coefficients in a homogeneous quantum gas. Our measurements provide parameter-free benchmarks for microscopic theories and establish an ultracold atom platform for studying spin caloritronics in strongly correlated matter.
\end{abstract}

\maketitle

Hydrodynamic transport of spin-1/2 fermions occurs in systems spanning many orders of magnitude in energy, from ultracold atomic gases to electron fluids in quantum materials to quark-gluon plasmas~\cite{OHaraScience,GrapheneHydro2018,NJPReview,BlochReview,UrbanReview}. Recent observations of spin polarization in heavy-ion collisions have stimulated the development of relativistic spin hydrodynamics~\cite{StarCollab,SpinHydroRelativistic,DissHydroSpinPol}. Spin-dependent transport underpins spintronic technologies, where ``spin caloritronic" properties~\cite{SpinCal2012,AdachiSpinSeebeck} are now widely studied, including the spin Seebeck effect~\cite{UchidaObsSeebeck,SpinThermoElec}, denoting the production of spin currents by temperature gradients, and the spin Peltier effect, denoting the generation of heat currents by spin-dependent forces.   Despite these advances, establishing quantitative tests of predictions of spin hydrodynamic and spin caloritronic properties remains challenging.

Unitary Fermi gases have been proposed for fundamental studies of spin caloritronics~\cite{KimHuseSpinDiff,DuineStoofSpinSeebeckFG}, motivated by initial measurements of spin-drag and spin-diffusivity in colliding clouds~\cite{SommerSpinDiff} more than a decade ago.  Unitary Fermi gases are strongly interacting and have universal properties~\cite{HoUniversalThermo}, dependent only on the density, temperature, and spin polarization, potentially enabling parameter-free tests of predictions. However, spatial inhomogeneity in the density has prevented direct determination of intrinsic spin transport coefficients and methods for measuring spin hydrodynamic and spin caloritronic properties in uniform quantum gases~\cite{NavonBox} have not been available.  

We report on the excitation and measurement of spin-dependent hydrodynamic modes in a spin-imbalanced
unitary Fermi gas of $^6$Li atoms, confined with uniform density in an optical box. 
By tracking the coupled evolution of the spin polarization and total density following the quench of a spatially periodic optical perturbation, we obtain direct experimental access to the fundamental energy and spin currents. We demonstrate that the initial evolution of the spin polarization is governed by the spin Seebeck current, enabling  independent determination of the universal spin hydrodynamic and spin caloritronic transport coefficients.

\subsection*{Probing Spin Transport via Spin Polarization Dynamics}

We investigate spin transport in the normal phase of a unitary Fermi gas of $^6$Li, comprising a mixture of the
two lowest hyperfine states, $1\equiv\,\uparrow$ and $2\equiv\,\downarrow$,  which serve as the spin components. The interactions are tuned to the broad $1-2$ Feshbach resonance~\cite{BartensteinFeshbach,JochimPreciseFeshbach}. Initially, a spin-balanced cloud is cooled by forced evaporation in a CO$_2$ laser trap~\cite{OHaraScience}. A spin-imbalanced mixture with a selected polarization is then prepared and transferred to an optical box formed by six sheets of repulsive light~\cite{LorinLinearHydro}, where it is allowed to reach equilibrium. For our experiments, the uniform total density is $n_0=n_{0\uparrow}+n_{0\downarrow}\simeq 3\times 10^{11}/{\rm cm}^3$  with a polarization $P_0=(n_{0\uparrow}-n_{0\downarrow})/n_0$ between $0$ and $0.40$.

\begin{figure}[htb]
\centering
\includegraphics[width=2.5in]{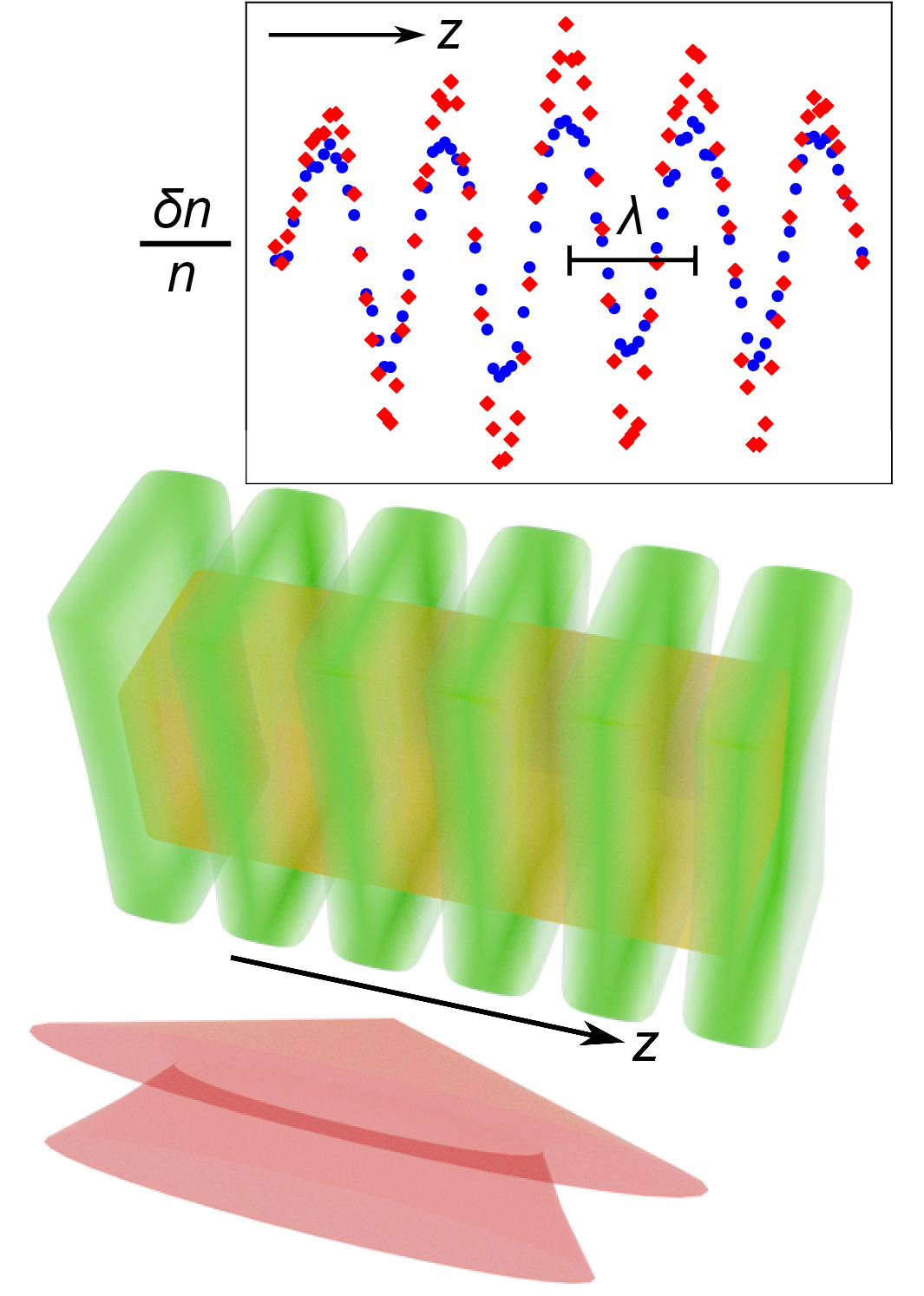}
    \caption{\textbf{Optical box with spatially periodic perturbation.} Atoms (yellow) are perturbed by a repulsive green light potential and imaged with resonant light (red arrow).  Inset (top) shows the differing equilibrium responses of the majority (blue) and minority (red) components, which sets the initial condition for measurement of the many-body spin hydrodynamics.}
    \label{fig:SetUp}
\end{figure}

Next, we smoothly ramp up a weak spatially periodic optical potential $\delta U$~\cite{XinHydroRelax}, made of repulsive 532 nm green light as shown in Fig.~\ref{fig:SetUp}, where $\delta U(z,t)=\delta U_0(t)(1+\cos qz)$ with wavevector $q = 2\pi/\lambda$ and wavelength $\lambda \simeq 33\,\mu$m. 
After equilibration, the resulting density modulations of the two spin components are measured by absorption imaging, Fig.~\ref{fig:SetUp}\,(inset). Two imaging pulses, each resonant with one spin component and separated in time by $10\,\mu$s, resolve the two components independently, on timescales short compared to the evolution time after $\delta U$ is quenched. Although the optical perturbation is the same for both species,  the normalized static response of the minority component $\delta n_\downarrow(z)/n_{0\downarrow}$ is larger than that of the majority component $\delta n_\uparrow(z)/n_{0\uparrow}$ for a spin-imbalanced mixture at temperatures $T<T_F$ with $T_F$ the Fermi temperature, while the normalized responses are identical in the Boltzmann limit~\cite{SupportOnline}.

\begin{figure}[htb]
\centering
\includegraphics[width=2.25in]{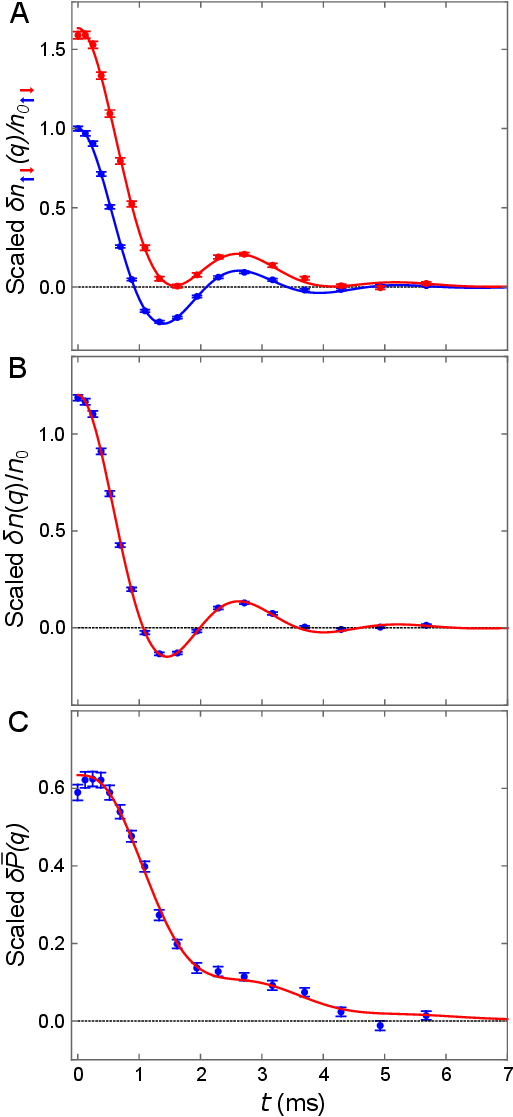}
    \caption{\textbf{Evolution of the spin polarization and spin density perturbations.} (\textbf{A}) Upper red curve: Minority spin Fourier component $\delta n_{\downarrow}(q,t)/n_{0\downarrow}$. Lower blue curve: Majority spin Fourier component $\delta n_{\uparrow}(q,t)/n_{0\uparrow}$. (\textbf{B}) Total density Fourier component $\delta n(q,t)/n_0$. (\textbf{C}) Spin polarization perturbation $\delta\bar{P}(q,t)\equiv\delta n_{\downarrow}/n_{0\downarrow}-\delta n_{\uparrow}/n_{0\uparrow}$.  Error bars denote the standard error for 40 shots.  Solid curves show the spin hydrodynamics model.}
    \label{fig:Spin12}
\end{figure}

Following an abrupt turn-off of the optical perturbation $\delta U(z,t)$  at time $t=0$,  we measure the amplitude of the dominant Fourier component $\delta n_{\uparrow\downarrow}(q,t)$  of the density changes $\delta n_{\uparrow\downarrow}(z,t)$ for each spin state.  
This is accomplished by a fast Fourier transform of the density profiles in the central region of the box, which is unaffected by reflections from the box edges. Fig.~\ref{fig:Spin12}A shows the scaled data in units of $\delta n_{\uparrow}(q,0)/n_{0\uparrow}\simeq 0.1$.  In Fig.~\ref{fig:Spin12}A, where $T=0.46\,T_F$ and $P_0=0.36$, the larger response of the minority component is observed as an obvious upward shift relative to that of the majority. Fig.~\ref{fig:Spin12}B shows the total density modulation $\delta n(q,t)/n_0$, where $\delta n(q,t)=\delta n_\uparrow(q,t)+\delta n_\downarrow(q,t)$.

Key to the measurements is the evolution of the polarization modulation $\delta P(q,t)$,  which provides a robust probe of the spin-transport properties. 
Experimentally, $\delta P(q,t)$ is obtained from the difference between the normalized density modulations of the two components. To first order in the density modulations,  $\delta P(q,t)=-(1-P_0^2)\,\delta\bar{P}(q,t)/2$, where $\delta\bar{P}\equiv\delta n_{\downarrow}/n_{0\downarrow}-\delta n_{\uparrow}/n_{0\uparrow}$, shown in Fig.~\ref{fig:Spin12}C, is the difference between the data curves of Fig.~\ref{fig:Spin12}A. In  Figs.~\ref{fig:Spin12}~(A-C), the solid curves show the fit of the spin hydrodynamics model, from which the spin and hydrodynamic transport coefficients are extracted, as discussed below.\\

\subsection*{Spin-Hydrodynamics Linear Response Model}

The measured relaxation of the density and polarization modulations provides access to the transport coefficients through a linear-response hydrodynamic model.
The model describes the coupled dynamics of the density, spin polarization, and temperature perturbations, $\delta n$, $\delta P$, and $\delta T$. To derive the evolution equations, we require explicit forms for the  energy, spin and heat current densities. 
For this purpose, we define the spin chemical potential  $h(P,n,T)\equiv(\mu_\uparrow-\mu_\downarrow)/2$. As the optical perturbation $\delta U(z)$ is the same for both spin states,  the change in the local chemical potentials $\mu_{\uparrow\downarrow}$ is identical, so the initial equilibrium satisfies $\delta h=0$ and $\delta T=0$.  In the evolution equations, just after $\delta U(z)$ is turned off, the initial conditions, $\delta\dot{n}_{\uparrow\downarrow}=0$, and hence $\delta\dot{P}=0$ and $\delta\dot{n}=0$, are then automatically satisfied by expressing the constitutive relations for the current densities in gradients of the $h$ and $T$ variables. 

Within linear irreversible thermodynamics, gradients of temperature and spin chemical potential drive coupled energy and spin currents.
The energy current density ${\mathbf J}_{\epsilon}$ and spin current density ${\mathbf J}_{\rm spin}={\mathbf J}_\uparrow -{\mathbf J}_\downarrow$  are expanded in  the thermodynamic forces  $\nabla (-1/T)$ and $\nabla (h/T)$ that enter the entropy production rate~\cite{SupportOnline,LandauFluids}, 
\begin{equation}
\left(
        \begin{array}{c}
          {\mathbf J}_{\epsilon} \\
          {\mathbf J}_{\rm spin} \\
        \end{array}
      \right)=-\left(
                \begin{array}{cc}
                  \kappa_\epsilon &  P_\epsilon \\
                 S_\epsilon & 2\sigma_s \\
                \end{array}
              \right)\left(
                       \begin{array}{c}
                        T^2 \nabla(-1/ T) \\
                        T \nabla (h/T) \\
                       \end{array}
                     \right)\,.
\label{eq:currentshT}
\end{equation}
Here, the extra factors of $T$ are chosen so that the transport coefficients have the usual dimensions. With $T^2 \nabla(-1/ T)=\nabla T$,  $\kappa_\epsilon$ is identified as the energy conductivity at constant $h/T$, which arises for adiabatic compression of a unitary gas, as in the first sound mode~\cite{SupportOnline}. Similarly, $S_\epsilon$ is the energy-spin Seebeck coefficient at constant $h/T$, which couples the spin and energy currents. 
At constant $T$, we have ${\mathbf J}_{\rm spin}=-\sigma_s\nabla(\mu_\uparrow-\mu_\downarrow)$,  where $\sigma_s$ is the spin conductivity. Combining this with ${\mathbf J}_{\rm spin}=-D_s\nabla(n_\uparrow-n_\downarrow)$ at constant $T$ and $n$ gives $n  D_s=2\sigma_s (\partial h/\partial P)_{Tn}$, which relates $\sigma_s$ to the spin diffusivity $D_s$. 

The Onsager reciprocal relation~\cite{SupportOnline,LandauFluids} requires  $P_\epsilon=T S_\epsilon$, where $P_\epsilon$ is the energy-spin Peltier coefficient. 
The constitutive relations  Eq.~\ref{eq:currentshT} can be related to the conventional description of spin caloritronic transport, where
the heat current density  ${\mathbf J}_{\rm heat}={\mathbf J}_{\epsilon}-h\,{\mathbf J}_{\rm spin}$ and the spin current density ${\mathbf J}_{\rm spin}$ are expanded in the $\nabla h-\nabla T$ representation. This yields the thermal conductivity $\kappa_T$, the heat-spin Seebeck coefficient $S_s=S_\epsilon-2\sigma_s\,h/T$, and the Peltier coefficient $P_s=TS_s$~\cite{SupportOnline}. The Kubo formula for $TS_s$~\cite{KimHuseSpinDiff} shows that $T S_\epsilon$ corresponds to the correlation between the energy and spin current densities.

Using the viscous damping force determined by the shear viscosity $\eta$~\cite{Bulk,ElliottScaleInv,SonBulkViscosity,StringariBulk} together with the constitutive relations Eq.~\ref{eq:currentshT} for the spin and energy current densities,  we can derive the evolution equations for the Fourier components of the perturbations to the density $\delta n(q,t)$, polarization $\delta P(q,t)$, and temperature $\delta T(q,t)$. These equations describe the dynamics  after $\delta U$ is extinguished at $t\equiv 0$.  For brevity, we discuss the evolution equation for $\delta P(q,t)$ here and defer detailed description of $\delta n(q,t)$, and $\delta T(q,t)$  to the supplementary material~\cite{SupportOnline}. To linear order in the perturbations, $\delta\dot{P}=-\nabla\cdot{\mathbf J}_{\rm spin}/n_0$. With Eq.~\ref{eq:currentshT},  we find
\begin{widetext}
\begin{equation}
\delta\dot{P}(q,t) =-D_s\,q^2\{\delta P(q,t)\!+c_{sn}[\delta\tilde{n}(q,t)-3/2\,\delta\tilde{T}(q,t)]\}
-S_\epsilon\,\frac{T_0}{n_0}\,q^2\,\delta\tilde{T}(q,t)\,,
\label{eq:deltaPdot}
\end{equation}
\end{widetext}
where $\delta\tilde{n}=\delta n/n_0$ and $\delta\tilde{T}=\delta T/T_0$. The first term describes spin diffusion, while the second term arises from the energy-spin Seebeck current.  The thermodynamic quantity $c_{sn}=-n_0\,(\partial P/\partial n)_{h,T}=-\delta P(q,0)/\delta\tilde{n}(q,0)$ is determined from the measured initial conditions~\cite{SupportOnline}. With $\delta T(q,0)=0$, Eq.~\ref{eq:deltaPdot} then assures that $\delta\dot{P}(q,0)=0$ as required.   

For static initial conditions, the predicted $\delta n(q,t)/n_0$ of Fig.~\ref{fig:Spin12}B and $\delta \bar{P}(q,t)$ of Fig.~\ref{fig:Spin12}C each contain contributions from a decaying oscillating first sound mode and two  decaying heat-spin diffusive modes. To fit the data,  we minimize the sum of the $\chi^2$ for $\delta n/n_0$ and $\delta\bar{P}$. We extract the transport properties in $\hbar/m$ units, $D_s\equiv\tilde{D}_s\,\hbar/m$,  $(T_0/n_0)\,S_\epsilon\equiv\tilde{S}_\epsilon\,\hbar/m$, $(n_0 m)^{-1}\eta\equiv\tilde{\eta}\,\hbar/m$, and $(n_0 k_B)^{-1}\kappa_\epsilon\equiv\tilde{\kappa}_\epsilon\,\hbar/m$, where $\tilde{D}_s$, $\tilde{S}_\epsilon$, $\tilde{\eta}$, and $\tilde{\kappa}_\epsilon$ are dimensionless fit parameters. For a unitary Fermi gas, these dimensionless parameters are universal functions of the reduced temperature $T/T_F$ and polarization $P$, enabling parameter-free comparisons of the measured values with predictions. 

In the hydrodynamic model, we employ a theoretical equation of state (EOS) for the pressure~\cite{SupportOnline,FrankLangZwergPhaseDiag}. 
We estimate the reduced temperature $T/T_F$ from the fitted frequency of the first sound mode $\omega_1=c_S\,q$, where the adiabatic sound velocity $c_S$ is obtained from the EOS. For our data, where $P\leq 0.36$, the reduced temperature obtained by this method is within 10\% of that found using the measured EOS for $P=0$\cite{KuThermo}. 

An important advantage of employing the constitutive relation for the energy current density, given by Eq.~\ref{eq:currentshT}, is that the first-sound diffusivity is independent of the spin transport coefficients  $S_\epsilon$ and $D_s$. 
In this representation, the first-sound diffusivity $D_1=\tilde{D}_1\hbar/m$~\cite{SupportOnline} is  
\begin{equation}
\tilde{D}_1=\frac{4}{3}\,\tilde{\eta}+\left(\frac{1}{\tilde{c}_{V_1}}-\frac{1}{\tilde{c}_{P_1}}\right)\,\tilde{\kappa}_\epsilon\,,
\label{eq:D1}
\end{equation}
where $\tilde{c}_{V_1}$ ($\tilde{c}_{P_1}$) is the heat capacity per particle at constant volume (pressure) in units of $k_B$.
For a spin-imbalanced unitary Fermi gas, the first-sound diffusivity has the usual~\cite{LandauFluids} form in terms of the energy conductivity $\kappa_\epsilon$ measured at constant $h/T$, which occurs for an adiabatic compression mode~\cite{SupportOnline}. Eq.~\ref{eq:D1} confirms the consistency of the evolution equations and simplifies the data analysis. 

In the following, we defer a discussion of $D_1$ to the supplementary material and focus on measurements of the spin transport properties $S_\epsilon$ and $D_s$ as functions of reduced temperature and polarization.

\subsection*{Extracting the Energy-Spin Seebeck Coefficient}

A central result of the hydrodynamic analysis is that the energy-spin Seebeck coefficient $\tilde{S}_\epsilon$ is completely determined by the leading non-vanishing time derivatives of the spin polarization and total density. 
Because the system starts in thermal, mechanical and diffusive equilibrium,  neither the  density nor the polarization initially evolve, so that  $\delta\dot n(q,0)=0$, $\delta\dot P(q,0)=0$. As all of the current densities initially vanish,  $\delta\dot T(q,0)=0$ as well. From Eq.~\ref{eq:deltaPdot}, these initial conditions require $\delta\ddot{P}(q,0)=0$. For the unitary gas, we find~\cite{SupportOnline},  $\delta \ddot{\tilde{T}}(q,0)=2/3\,\delta\ddot{\tilde{n}}(q,0)$, and Eq.~\ref{eq:deltaPdot} yields
\begin{equation}
\delta\dddot P(q,0)=-\frac{2}{3}\frac{\hbar\, q^2}{m}\,\delta\ddot{\tilde{n}}(q,0)\,\tilde{S}_\epsilon\,.
\label{eq:Ptripledot}
\end{equation}
Here, $\delta\ddot{\tilde{n}}(q,0)=\delta\ddot{n}(q,0)/n_0$ is well-determined from the short-time evolution of the total density perturbation, shown in Fig.~\ref{fig:Spin12}B. 
Physically, at short times, the temperature change arises from adiabatic compression/expansion of the total density at constant polarization, which keeps  $h/T$ constant in a unitary Fermi gas~\cite{SupportOnline}.  Hence, the energy-spin Seebeck current at constant $h/T$ in Eq.~\ref{eq:currentshT}  governs the short-time evolution of the polarization.

\begin{figure}
\centering
\includegraphics[width=3.0in]{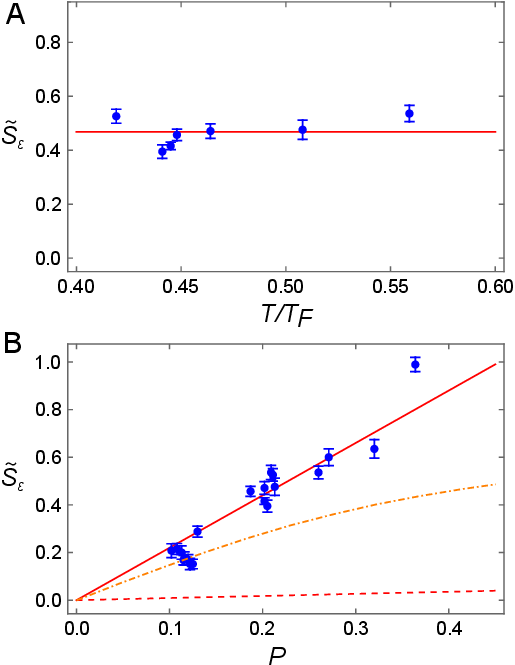}
\caption{\textbf{Energy-spin Seebeck coefficient $S_\epsilon=\tilde{S}_\epsilon\,\frac{n}{T}\,\frac{\hbar}{m}$.} (\textbf{A}) $\tilde{S}_\epsilon$ versus reduced temperature for $P \simeq  0.2$.  Red line: $\tilde{S}_\epsilon=0.468$. (\textbf{B}) $\tilde{S}_\epsilon$ versus polarization. Red line: Linear fit $\tilde{S}_\epsilon=2.20\,P$. Red dashed line: Kinetic theory limit for $T/T_F=0.45$.  Orange dot-dashed curve:  Quantum Boltzmann theory prediction for $T/T_F=0.45$~\cite{SupportOnline}. Error bars from the $\chi^2$ fits of the spin hydrodynamics model.}
\label{fig:Seps}
\end{figure} 

Remarkably, $\tilde{S}_\epsilon$ determines the short-time evolution of $\delta P(q,t)$,
enabling measurement of the fundamental energy-spin Seebeck coefficient from the fits of the model to the relaxation data.
Fig.~\ref{fig:Seps}A shows the dimensionless energy-spin Seebeck coefficient $\tilde{S}_\epsilon$ versus reduced temperature $T/T_F$ for fixed polarization in the range $0.187\leq P \leq 0.213$, with an average $\langle P\rangle=0.20$. Over the measured temperature range, $\tilde{S}_\epsilon$ is approximately independent of $T/T_F$. 
The red line shows the fit $\tilde{S}_\epsilon=0.468$. 

Fig.~\ref{fig:Seps}B shows the energy-spin Seebeck coefficient as a function of $P$. Since the temperature dependence is negligible over the range shown in Fig.~\ref{fig:Seps}A, we combine data for $0.38<T/T_F<0.58$, with an average $\langle T/T_F\rangle = 0.47$. 
$\tilde{S}_\epsilon$ appears to vary linearly with $P$, in qualitative agreement with predictions for a unitary Fermi gas in the two-body kinetic theory limit~\cite{SupportOnline,KimHuseSpinDiff}. The positive sign for a polarization $P>0$, corresponds to a positive correlation between the energy and spin currents. In a kinetic theory picture~\cite{KimHuseSpinDiff} of a unitary gas with a temperature gradient, a positive sign arises from the inverse scaling of the collision rate with relative speed, which causes a minority spin-down atom in a majority spin-up bath to be preferentially kicked toward the higher temperature side, producing a spin current in the same direction as the energy current.  For the two-body kinetic theory (KT) limit, we find $\tilde{S}_{\epsilon KT}\simeq 0.26\, P\,\tilde{D}_{sKT}$,  where $\tilde{D}_{sKT}\simeq 1.1 (T/T_F)^{3/2}$ is the diffusivity~\cite{SupportOnline,BruunSpinDiff}. For $T/T_F=0.45$,  $\tilde{S}_{\epsilon KT}=0.088\,P$,  which is shown as the red dashed line.  

The red line in Fig.~\ref{fig:Seps}B shows a linear $\chi^2$ fit to the data, $\tilde{S}_\epsilon=2.20(06)\,P$. For comparison~\cite{SupportOnline}, an independent extraction based on Eq.~\ref{eq:Ptripledot}  gives a consistent  slope of $1.97$. The orange dot-dashed curve shows a quantum Boltzmann prediction for $T_F=0.45$ including medium effects~\cite{SupportOnline}, which have been shown in previous studies to be quantitatively important in the considered quantum critical regime~\cite{Enss2012}. For $P\leq 0.20$, this theory yields $\tilde{S}_\epsilon\simeq 1.44\,P$.   The observed sign and magnitude are consistent with the quantum Boltzmann prediction.   
As the determination of $\tilde{S}_\epsilon$  is independent of the equation of state, which enters only through the temperature calibration, the measured coefficient provides a direct benchmark for microscopic theories.

\subsection*{Extracting the Spin Diffusivity}

\begin{figure}
\centering
\includegraphics[width=2.9in]{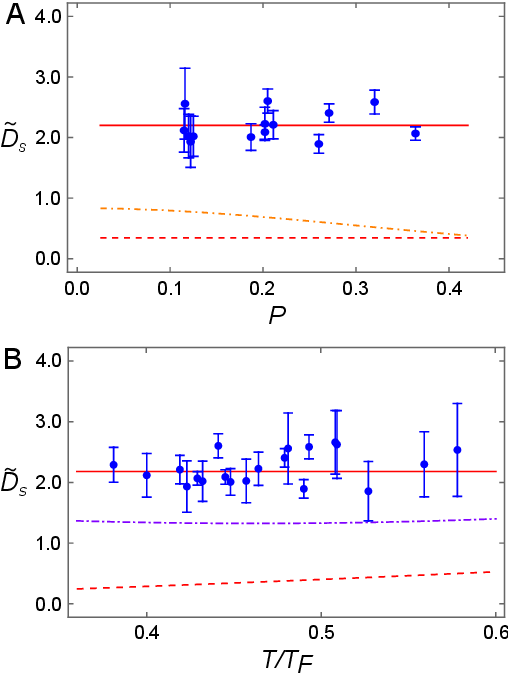}
    \caption{\textbf{Spin diffusivity $D_s=\tilde{D}_s\,\frac{\hbar}{m}$.} (\textbf{A}) $\tilde{D}_s$ versus polarization. Red line: $\tilde{D}_s=2.20$. Red dashed line: Kinetic theory prediction for $T/T_F=0.45$.  Orange dot-dashed curve: Quantum Boltzmann theory prediction for $T/T_F=0.45$. (\textbf{B}) $\tilde{D}_s$ versus reduced temperature. Red line: $\tilde{D}_s=2.20$. Red dashed line: Kinetic theory limit $\tilde{D}_{sKT}\simeq 1.1 (T/T_F)^{3/2}$. Purple dot-dashed curve: Prediction~\cite{EnssSpinDiff}. Error bars from the $\chi^2$ fits of the spin hydrodynamics model.}
    \label{fig:Ds}
\end{figure}

The homogeneous box geometry enables a direct measurement of the intrinsic spin diffusivity of a unitary Fermi gas without averaging over density inhomogeneities.
We determine the spin diffusivity from the measured decay of the polarization $\delta\bar{P}(q,t)$ in Fig.~\ref{fig:Spin12}C. 
Fig.~\ref{fig:Ds}A shows the spin diffusivity $\tilde{D}_s$ obtained from the fits for $0.4\leq T/T_F\leq 0.5$ as a function of polarization $P$.  
The measured spin diffusion coefficients $\tilde{D}_s$ have a negligible dependence on the polarization, in qualitative agreement with  kinetic theory predictions~\cite{KimHuseSpinDiff,SupportOnline}, shown as a red dashed line for $T/T_F=0.45$. The orange dot-dashed curve shows the quantum Boltzmann equation prediction~\cite{SupportOnline}, which is smaller in magnitude and exhibits a stronger polarization dependence than observed, suggesting that improved calculations are needed.

Fig.~\ref{fig:Ds}B shows  the spin diffusivity $\tilde{D}_s$  obtained from the fits for $0.1\leq P\leq 0.4$ versus $T/T_F$, where we assume a negligible dependence on $P$, as shown in Fig.~\ref{fig:Ds}A. The red dashed curve shows the kinetic theory limit $\tilde{D}_{sKT}\simeq 1.1 (T/T_F)^{3/2}$~\cite{SupportOnline,BruunSpinDiff}. The red line shows the fit $\tilde{D}_s=2.20$. The purple dot-dashed curve is the prediction~\cite{EnssSpinDiff} using the strong-coupling Luttinger-Ward theory, where $\tilde{D}_s\simeq 1.4$ over the measured temperature range. The measured $\tilde{D}_s$ is nearly independent of reduced temperature for $T/T_F=0.4-0.6$, consistent with predictions~\cite{EnssSpinDiff} and with previous measurements in cigar-shaped samples~\cite{SommerSpinDiff}.

The fitted value $\tilde{D}_s=2.20$ is much smaller than obtained in Ref.~\cite{SommerSpinDiff}, where $\tilde{D}_s\simeq 6.3$ in the same temperature range. There, the vanishing density near the cloud edges substantially increases the extracted values~\cite{SommerSpinDiff,SchaferSpinDiff}. 
$\tilde{D}_s=2.20$ is closer to predictions for a Pauli-blocked degenerate sample~\cite{BruunSpinDiff}. However, the data show no indication of the $(T_F/T)^2$ divergence at low temperature, which characterizes Pauli-blocking. Remarkably, $\tilde{D}_s=2.20$ is comparable to the transverse diffusivity predicted~\cite{EnssTransvSpin} and measured~\cite{ThywissenTransvSpin} for a coherently prepared unitary Fermi gas.

\subsection*{Discussion and Conclusions}

We have shown that the short time dynamics of spatially periodic perturbations in the spin polarization
and total density reveals the fundamental spin-calorimetric properties of a unitary Fermi gas in a homogeneous box.
By fitting the full spin-hydrodynamic evolution, we obtain the energy-spin Seebeck coefficient, spin diffusivity, and first-sound diffusivity, which characterize coupled spin, heat, and momentum transport in a strongly interacting quantum fluid. These measurements establish a quantitative framework for studying universal spin hydrodynamics and open the way to systematic investigations of spin-caloritronic transport in strongly correlated quantum matter.

$^*$Corresponding author: jethoma7@ncsu.edu


%

\section*{Acknowledgments}
We thank Thomas Sch\"{a}fer  for stimulating conversations and Tilman Enss for helpful discussion of his previous work. \\
\vspace*{-0.1in}\paragraph*{Funding:}
The work of D.~Z. and J.~E.~T. was supported by the U. S. National Science Foundation (PHY-2307107).
The work of J.~L. was supported by the Deutsche Forschungsgemeinschaft 
(DFG, German Research Foundation) under Germany’s Excellence Strategy 
Cluster of Excellence Matter and Light for Quantum Computing (ML4Q) 
EXC 2004/1 390534769, and by the DFG Collaborative Research Center 
(CRC) 183 Project No. 277101999.\\
\vspace*{-0.1in}\paragraph*{Author contributions:}
D.~Z. performed the experimental measurements. D.~Z. and J.~E.~T. developed the linear response model and analyzed the data.
J. L. provided the theoretical equation of state and predicted the transport properties. 
All authors contributed to the interpretation of the data and the preparation of the manuscript.

\newpage
\widetext
\setcounter{figure}{0}
\setcounter{equation}{0}
\renewcommand{\thefigure}{S\arabic{figure}}
\renewcommand{\theequation}{S\arabic{equation}}

\section{Experimental Methods}
\label{sec:expt}

This section describes the experimental methods used to measure the density perturbations for the majority and minority spin components, which determine the perturbations for the total density and polarization shown in the main text. 
In the experiments, we create a uniform, spin-imbalanced unitary $^6$Li Fermi gas in a 532 nm repulsive optical box potential $U_0$  projected by two micromirror arrays. The box dimensions are $150\,\mu {\rm m}\times50\,\mu {\rm m}\times 50\,\mu {\rm m}$, containing initially uniform densities for the majority $n_\uparrow$ and minority $n_\downarrow$ spin components in static equilibrium.  The corresponding uniform polarization is
\begin{equation}
P=\frac{n_\uparrow-n_\downarrow}{n_\uparrow+n_\downarrow}\equiv\frac{\sigma}{n},
\label{eq:Polariz}
\end{equation}
where $\sigma = n_\uparrow-n_\downarrow$ is the magnetization density and $n=n_\uparrow+n_\downarrow$ is the total density.

After creating the sample, we smoothly apply a repulsive spatially periodic 532 nm optical potential of the form $\delta U(z,t)=\delta U_0(t)\,[1+\cos(qz)]$, where $q=2\pi/\lambda$ with $\lambda \simeq 33\,\mu$m. This optical potential creates a  density perturbation $\delta n_{\uparrow\downarrow}(z)$ for each species, which is allowed to reach static equilibrium with a uniform temperature $T=T_0$. The initial density profiles $n_{\uparrow\downarrow}(z,0)=n_{0\uparrow\downarrow}+\delta n_{\uparrow\downarrow}(z,0)$ are determined by the local chemical potentials $\mu_{\uparrow\downarrow}$,
\begin{equation}
\mu_{\uparrow\downarrow}[n_\uparrow(z,0),n_\downarrow(z,0),T_0]+U_0+\delta U(z,0)=\mu_{g\uparrow\downarrow}\,.
\label{eq:mu}
\end{equation}
Here, both $U_0$ and the optical perturbation at 532 nm are identical for both species, due to the large detuning. 
We perform a spatial Fourier transform of the density profiles to extract the dominant Fourier component $\delta n_{\uparrow\downarrow}(q,0)$ for each species.

After the $\delta n_{\uparrow\downarrow}(q,0)$ are in static equilibrium, $\delta U$ is abruptly extinguished, at time $t\equiv 0$. The resulting $\delta n_{\uparrow\downarrow}(q,t)/n_{0\uparrow\downarrow}$ are measured by absorption imaging, using two imaging pulses, each resonant with one spin component, and separated in time by $10\,\mu$s, short compared to the time scale for the evolution.  To suppress small systematic errors arising from the first image affecting the second, the imaging sequence is done in both orders and averaged. The change is found to be negligible compared to the average.

Since the global chemical potentials $\mu_{g\uparrow\downarrow}$ are spatially constant, and $U_0$ and $\delta U(z,0) $ are the same for both species,  the spin chemical potential
\begin{equation}
h\equiv\frac{\mu_\uparrow-\mu_\downarrow}{2}
\label{eq:h}
\end{equation}
is spatially constant in equilibrium, $h=h_0$. 
Taking $h=h(n_\uparrow,n_\downarrow,T)$, our initial conditions, $\delta h\equiv h-h_0=0$ and $\delta T\equiv T-T_0=0$, require $\delta n_\downarrow(0)/\delta n_\uparrow(0)=-(\partial h/\partial n_\uparrow)_T/(\partial h/\partial n_\downarrow)_T$.

In practice, we measure $\delta n_{\uparrow\downarrow}(q,t)/n_{0\uparrow\downarrow}$ to reduce fluctuations arising from variations in the atom numbers. Using Eq.~\ref{eq:Polariz}, we define $P=P_0+\delta P$, where the measurements determine $\delta P$  to first order in the density  perturbations,
\begin{equation}
\delta P = \frac{1-P_0^2}{2}\left(\frac{\delta n_\uparrow}{n_{0\uparrow}}-\frac{\delta n_\downarrow}{n_{0\downarrow}}\right)\,.
\label{eq:1.14}
\end{equation}
The measured time dependence of $\delta P(q,t)$ provides a new method of measuring spin transport properties in spin-imbalanced gases.

We also measure $\delta\tilde{n}(q,t)$, where $\delta\tilde{n}\equiv\delta n/n_0= (\delta n_\uparrow+\delta n_\downarrow)/n_0$, with $n_0$ the spatially constant total density,
\begin{equation}
\delta\tilde{n}=\frac{1+P_0}{2}\,\frac{\delta n_\uparrow}{n_{0\uparrow}}+\frac{1-P_0}{2}\,\frac{\delta n_\downarrow}{n_{0\downarrow}}\,.
\label{eq:1.13}
\end{equation}
As discussed below,   $\delta\tilde{n}(q,t)$ and $\delta P(q,t)$ each contain contributions from a decaying oscillating first sound mode and two coupled decaying heat-spin diffusive modes. 

For thermometry, we employ an in-situ method for measuring the reduced temperature $\theta=T/T_F$ (see Eq.~\ref{eq:FermiEn}), which is determined by the fitted frequency $\omega_1$ of the first sound mode. In the long wavelength limit, $\omega_1=c_Sq$, where $mc_S^2=(\partial p/\partial n)_{P\theta}$ determines the adiabatic sound speed $c_S$ from the theoretical equation of state for the pressure $p$ (see Eq.~\ref{eq:4.1S}). 

\begin{figure}
\centering
\includegraphics[width=3.5in]{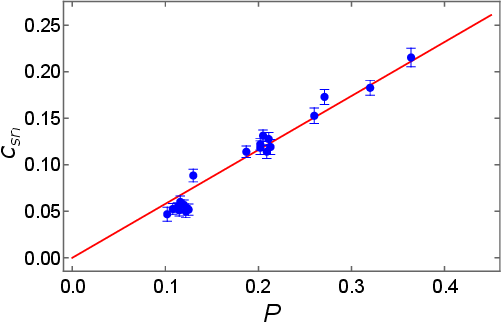}
 \caption{\textbf{Spin-density coupling  coefficient $c_{sn}$ versus polarization.}
    Red dots: Measured ratio $c_{sn}$ for $T/T_F=0.38-0.58$. Red Line: Linear fit $c_{sn}(P)=0.58\,P$. Error bars denote the standard error for 40 shots. }
    \label{fig:CsnvsP}
\end{figure}  

The ratio $c_{sn}\equiv -\delta P(q,0)/\delta\tilde{n}(q,0)$ plays a central role in the linear response model. As discussed in \S~\ref{sec:natvar}, $c_{sn}$ is a thermodynamic quantity that is measured from the initial conditions in the experiments and can be compared to predictions based on the theoretical equation of state. For our experiments, where $n_\uparrow>n_\downarrow$, i.e., spin $\downarrow$ is the minority species, we find $c_{sn}>0$. At uniform temperature, with $h=h(P,n,T)$, $\delta h =0$ requires
\begin{equation}
c_{sn}\equiv -\frac{\delta P(q,0)}{\delta\tilde{n}(q,0)}=-n_0\left(\frac{\partial P}{\partial n} \right)_{\!\!hT}=n_0\,\frac{(\partial h/\partial n)_{PT}}{(\partial h/\partial P)_{nT}}\,.
\label{eq:1.15}
\end{equation}
Eq.~\ref{eq:1.15} shows that $\delta P(q,0)\neq 0$ for $(\partial h/\partial n)_{P,T}\neq 0$, enabling spin transport measurements based on the polarization perturbation. As $c_{sn}$ is a thermodynamic quantity, the measured value is independent of the wavevector $q$. Fig.~\ref{fig:CsnvsP} shows the measured $c_{sn}$ versus polarization $P$ for fitted reduced temperatures in the range $0.38<T/T_F<0.58$, where the average $\langle T/T_F\rangle =0.47$.  A linear fit to all of the data gives $c_{sn}(P)=0.58\,P$ (red line). Eq.~\ref{eq:csnS}, discussed below, gives $c_{sn}$ in terms of the predicted $h(P,n,T)$ of Eq.~\ref{eq:htheor}. For $T/T_F=0.45$, the predicted $c_{sn}(P)=0.544\,P$, just 7\% lower than the measured value, confirming that the theoretical equation of state (see \S~\ref{sec:thermo})  is in close agreement with the measurements.

In the high temperature limit, $h=\frac{k_BT}{2}\,\ln (n_\uparrow/n_\downarrow)=\frac{k_BT}{2}\,\ln [(1+P)/(1-P)]$ is independent of $n$, so that $(\partial h/\partial n)_{PT}\rightarrow 0$ and $\delta P\rightarrow 0$, i.e.,  $\delta n_\downarrow(0)/\delta n_\uparrow(0)=-(\partial h/\partial n_\uparrow)_T/(\partial h/\partial n_\downarrow)_T=n_{0\downarrow}/n_{0\uparrow}$, so the normalized responses are identical. Hence, spin transport cannot be measured by relaxation of the Fourier components of the polarization in the classical Boltzmann regime.

\section{Supplemental Material}
\label{sec:supplement}
The supplemental text is organized as follows. 
In \S~\ref{sec:thermo}, we discuss the theoretical equation of state, which is used to determine the reduced temperature and to determine the thermodynamic properties. We derive the thermodynamic relations for the changes in pressure $\delta p$, density $\delta n$, polarization $\delta P$, and temperature $\delta T$, which are needed to develop a spin hydrodynamics linear response model. Key to the model are the consistent definitions of the heat, energy, and spin current densities, and the choice of the fundamental transport coefficients, which are described in detail in \S~\ref{sec:HeatSpin}. To provide an initial comparison of the measurements to predictions, the spin transport coefficients are determined using a quantum Boltzmann equation treatment with medium corrections, discussed briefly in \S~\ref{sec:Boltzmann}.   In \S~\ref{sec:Hydro}, we derive the evolution equations for $\delta n$, $\delta P$, and $\delta T$, which are used to model the data. We show that the chosen set of transport coefficients imposes important constraints for the evolution equations, enabling robust determination of the energy-spin Seebeck coefficient, the spin diffusivity, and the first sound diffusivity in a spin-imbalanced unitary Fermi gas.

\subsection{Thermodynamics}
\label{sec:thermo}
The spin hydrodynamics model developed below requires thermodynamic relations connecting the experimentally accessible variables, total density and polarization, to the temperature,   pressure, and chemical potentials of the unitary gas. In this section, we first summarize the equation of state obtained from a self-consistent Luttinger-Ward T-matrix approach, which provides the universal thermodynamic functions entering the analysis. We then derive the linear thermodynamic relations that express the pressure and temperature perturbations in terms of the density and polarization perturbations. These relations determine all thermodynamic coefficients appearing in the hydrodynamic evolution equations derived in \S~\ref{sec:Hydro}.

For the following discussion, we define the local Fermi energy scale $\epsilon_F$ and the
corresponding Fermi temperature scale $T_F$ for atoms of mass $m$ by
\begin{equation}
k_BT_F(n)=\epsilon_F(n)\equiv\frac{\hbar^2}{2m}(3\pi^2 n)^{2/3}\equiv\frac{m}{2}\,{\rm v}_F^2(n)\,.
\label{eq:FermiEn}
\end{equation}
Here, $\epsilon_F(n)$ is the local Fermi energy for an ideal gas of total density $n$ in a  $50-50$ mixture of two spin states, 
and  ${\rm v}_F(n)$ is the corresponding Fermi speed. 

The pressure  $p(P,n,T)$, for a spin-imbalanced unitary Fermi gas can be written in the universal form,
\begin{equation}
p(P,n,T)=n\epsilon_F(n)\,f_p(P,\theta)\,,
\label{eq:4.1S}
\end{equation}
where  $\theta=T/T_F(n)$ is the reduced temperature and $f_p(P,\theta)$ is a dimensionless universal pressure function. 

Similarly, we define the spin chemical potential,
\begin{equation}
h(P,n,T)=\frac{\mu_\uparrow-\mu_\downarrow}{2}=\epsilon_F(n)\,f_h(P,\theta)\,,
\label{eq:htheor}
\end{equation}
where $f_h(P,\theta)$ is a dimensionless universal scaling function. 

Finally, the entropy per particle, in units of $k_B$, takes the form
\begin{equation}
s_1(P,n,T)=k_Bf_{s_1}(P,\theta),
\label{eq:5.1S1}
\end{equation}
where $f_{s_1}(P,\theta)$ is a dimensionless universal function.

The scaling functions $f_p, f_h$ and $f_{s_1}$ are obtained from a 
self-consistent $T$-matrix approximation of the unitary Fermi gas at 
finite spin imbalance \cite{FrankLangZwergPhaseDiag}. The method is conserving in 
the Baym--Kadanoff sense, \emph{i.e.}, all thermodynamic relations are 
fulfilled. It is based on the Luttinger--Ward functional 
\cite{Luttinger1960,Haussmann2007}, which yields the grand potential 
$\Omega(T,\mu,h,1/a,V)=-pV$.
Formally, the grand potential is given by
\begin{align}\label{eq:LW_functional}
\Omega[G]=k_B T \left(\mathrm{Tr}\{\ln G + [1-G_0^{-1}G]\}+\Phi[G]\right),
\end{align}
where $G=\{G_\uparrow,G_\downarrow\}$ denotes the imaginary-time single-particle 
Green's functions
\begin{align}
G_i(\mathbf{x},\tau)=-\langle 
\mathcal{T}\,\hat\psi_i(\mathbf{x},\tau) 
\hat\psi^\dagger_i(\mathbf{0},0)\rangle,
\end{align}
and $\mathcal{T}$ is the time-ordering operator. From it the density 
is obtained as
\begin{align}
n_i=G_i(\mathbf{x}=\mathbf{0},\tau=0^-)\,.
\end{align}

Equation~\eqref{eq:LW_functional} is formally exact provided that 
$\Phi$ contains the sum of all topologically allowed two-particle 
irreducible skeleton diagrams.
Thermodynamic consistency requires that $G_i$ satisfies the Dyson 
equation
\begin{align}\label{eq:Dyson}
G_i^{-1}(\mathbf{k},\omega_j)=G_{0,i}^{-1}(\mathbf{k},\omega_j)-\Sigma_i[G],
\end{align}
with bare propagator
\begin{align}
G_{0,i}(\mathbf{k},\omega_j)=\frac{1}{i\hbar\omega_j-\epsilon_k+\mu_i},
\end{align}
where $\omega_j=(2j+1)\pi k_B T/\hbar$ are fermionic Matsubara 
frequencies and $\epsilon_k=\hbar^2 k^2/(2m)$.
The self-energy is given by
\begin{align}
\Sigma_i[G]=\frac{\delta \Phi[G]}{\delta G_i}.
\end{align}

In practice, $\Phi$ is truncated to all particle--particle ladder 
diagrams, yielding the self-consistent $T$-matrix approximation. This 
truncation preserves thermodynamic consistency, but no longer 
constitutes an exact description of the Fermi gas. For the balanced 
system, it yields thermodynamic quantities in excellent quantitative 
agreement with experiment, e.g., $T_c/T_F\simeq 0.16$ and the Bertsch 
parameter $\xi_s\simeq 0.36$ at $T=0$~\cite{Haussmann2007}.

Within this approximation, the self-energy reads
\begin{align}\label{eq:Sigma}
\Sigma_i(\mathbf{k},\omega_j)=\int \frac{d^3 Q}{(2\pi)^3}\,k_B 
T\sum_l\Gamma(\mathbf{Q},\Omega_l)G_{\bar i}(\mathbf{Q}-\mathbf{k},\Omega_l-\omega_j),
\end{align}
where $\Omega_l=l\,2\pi k_B T/\hbar$ are bosonic Matsubara frequencies 
and $\bar i$ denotes spin opposite to $i$.
The vertex function
\begin{align}\label{eq:Gamma}
\Gamma(\mathbf{Q},\Omega_l)=\frac{1}{1/g + M(\mathbf{Q},\Omega_l)}
\end{align}
resums the geometric series of particle--particle ladders with the pair propagator
\begin{align}\label{eq:M}
M(\mathbf{Q},\Omega_l)=\int\frac{d^3 k}{(2\pi)^3}\left[k_B 
T\sum_j G_\uparrow(\mathbf{k},\omega_j)G_\downarrow(\mathbf{Q}-\mathbf{k},\Omega_l-\omega_j)-\frac{1}{2\epsilon_k}\right]
\end{align}
in Eq.~\ref{eq:Gamma}, where $g=4\pi\hbar^2 a/m$ denotes the coupling constant with the s-wave scattering length $a\rightarrow\infty$ at unitarity.

The leading instability of the unitary Fermi gas is the transition to 
a superfluid, signaled by a divergence of $\Gamma(\mathbf{0},0)$ 
(Thouless criterion). Close to this instability, straightforward 
fixed-$g$ iterations converge poorly. We therefore dynamically adjust 
the bare coupling during the iterative solution of 
Eqs.~\eqref{eq:Dyson}--\eqref{eq:M} such that $\Gamma(\mathbf{0},0)$ 
remains fixed. This significantly stabilizes convergence at low 
temperature. As a consequence, the code converges at values of $g$ 
that are not fixed a priori.

All calculations are performed in units with $k_B T=1$, such that the 
thermodynamics is obtained as a function of the dimensionless 
variables $\mu_i/(k_BT)$ and $g\sqrt{m^3 k_B T/\hbar^6}$.
The grand potential directly yields $f_p$, and the difference of 
chemical potentials $(\mu_\uparrow-\mu_\downarrow)/2$  gives $f_h$. The entropy density can be written as
\begin{align}
s=n s_1=(p+{\cal E}-\sum_i \mu_i\, n_i )/T\,,
\end{align}
where the internal energy density can be written as \cite{Frank2019}
\begin{align}
{\cal E}=\sum_i \int \frac{d^3 k}{(2\pi)^3} \epsilon_k 
G_i(\mathbf{k},\tau=0^-)+\frac{k_B 
T}{2}\sum_{i,j}\Sigma_i(\mathbf{k},\omega_j)G_i(\mathbf{k},\omega_j)
\end{align}
and $p=2 {\cal E}/3$.
However, because the iteration keeps $\mu_i/(k_BT)$ fixed rather 
than the polarization $P$,
obtaining results at fixed polarization and at unitarity requires a 
two-dimensional Newton search in $(\mu/(k_BT), h/(k_BT))$.
This procedure enforces simultaneously the desired total density and 
polarization, allowing us to express the universal scaling functions 
$f_p$, $f_h$, and $f_{s_1}$ in terms of the physical variables 
$(P,\theta)$ with $\theta=T/T_F$.

To derive the evolution equation for $\delta n$, we require the first order pressure change, $\delta p$,
\begin{equation}
\delta p=\left(\frac{\partial p}{\partial n}\right)_{\!\!PT}\!\!\delta n+\left(\frac{\partial p}{\partial T}\right)_{\!\!Pn}\!\!\delta T+\left(\frac{\partial p}{\partial P}\right)_{\!\!nT}\!\!\delta P,
\label{eq:4.2S}
\end{equation}
where $\delta n \equiv n-n_0$, $\delta T\equiv T-T_0$ and $\delta P \equiv P-P_0$ are the perturbations in the density, temperature, and polarization, relative to their spatially constant background values, $n_0$, $T_0$, and $P_0$.

In Eq.~\ref{eq:4.2S}, the first two terms are evaluated at constant polarization, $P$. The first term is the isothermal sound speed $c_T$  at constant $P$,
\begin{equation}
m\,c_T^2=\left(\frac{\partial p}{\partial n}\right)_{\!\!PT}.
\label{eq:4.3S}
\end{equation}

It is convenient to factor out $mc_T^2$ in the second term of Eq.~\ref{eq:4.2S} by using the chain rule and Eq.~\ref{eq:4.3S},
\begin{equation}
\left(\frac{\partial p}{\partial T}\right)_{\!\!Pn}=-\left(\frac{\partial p}{\partial n}\right)_{\!\!PT}\left(\frac{\partial n}{\partial T}\right)_{\!\!Pp}=mc_T^2\, n_0\,\beta_e\,.
\label{eq:1.1b}
\end{equation}
Here,  the expansivity $\beta_e$  is evaluated at constant polarization $P$.  With the volume per particle $V_1=1/n$,
\begin{equation}
\beta_e\equiv\frac{1}{V_1}\left(\frac{\partial V_1}{\partial T}\right)_{Pp}=-\frac{1}{n}\left(\frac{\partial n}{\partial T}\right)_{Pp},
\label{eq:beta}
\end{equation}
which has a dimension of inverse temperature. For the unitary gas,  $\beta_e$ has a simple relation to  the Landau-Placek parameter $\epsilon_{LP}=c_{P_1}/c_{V_1}-1$, which is also evaluated at constant $P$, 
\begin{equation}
\beta_e\, T=\frac{3}{2}\,\epsilon_{LP}\,,
\label{eq:beta}
\end{equation}
Eq.~\ref{eq:beta} follows from the univeral pressure relation for a unitary Fermi gas $p=2/3\,{\cal E}$~\cite{HoUniversalThermo} and
$(\partial{\cal E}/\partial T)_{Pn}=(\partial{\cal E}/\partial T)_{n_\uparrow,n_\downarrow}= n\,c_{V_1}$. 
Then $(\partial p/\partial T)_{Pn}=2/3\,n\,c_{V_1}$ in Eq.~\ref{eq:1.1b}, yielding $\beta_e\, m\,c_T^2 = 2/3\,c_{V_1}$. Inserting this in the well-known relation $c_{\!P_1}-c_{V_1}=mc_T^2\,\beta_e^2\,T$, where $c_{\!P_1}$ is the heat capacity per particle at constant $P$, we obtain Eq.~\ref{eq:beta}. 

The last term in Eq.~\ref{eq:4.2S} is directly evaluated using Eq.~\ref{eq:4.1S}, yielding
\begin{equation}
\left(\frac{\partial p}{\partial P}\right)_{\!\!nT}=\left(\frac{\partial p}{\partial P}\right)_{\!\!n\theta}=n_0\epsilon_F\left(\frac{\partial f_p}{\partial P}\right)_{\!\!\theta}\,.
\label{eq:4.6S}
\end{equation}

With Eqs.~\ref{eq:4.3S},~\ref{eq:1.1b},and~\ref{eq:4.6S}, Eq.~\ref{eq:4.2S} takes the form
\begin{equation}
\delta p=n_0\,m c_T^2\,(\,\delta\tilde{n}+3/2\,\epsilon_{LP}\,\delta\tilde{T}\,) + n_0\epsilon_F\left(\frac{\partial f_p}{\partial P}\right)_{\!\!\theta}\delta P,
\label{eq:deltap}
\end{equation}
where we have defined the dimensionless density and temperature changes that we will use in the evolution equations,
\begin{equation}
\delta\tilde{n}\equiv\frac{\delta{n}}{n_0}\hspace{0.25in}\delta\tilde{T}\equiv\frac{\delta{T}}{T_0}\,.
\label{eq:1.2Sb}
\end{equation}

The first order temperature change, $\delta T$, can be directly obtained from entropy per particle $s_1(P,\theta)$ 
of the unitary Fermi gas, Eq.~\ref{eq:5.1S1},
\begin{equation}
\delta s_1=\left(\frac{\partial s_1}{\partial P}\right)_{\!\!\theta}\!\delta P+\left(\frac{\partial s_1}{\partial\theta}\right)_{\!\!P}\left[\left(\frac{\partial \theta}{\partial T}\right)_{\!\!n}\!\!\delta T+\left(\frac{\partial \theta}{\partial n}\right)_{\!\!T}\!\!\delta n\right]\,.
\label{eq:1.1s1}
\end{equation}

We can evaluate the last two terms of Eq.~\ref{eq:1.1s1} using $(\partial\theta/\partial T)_n =\theta/T_0$, $(\partial\theta/\partial n)_T=-2/3\,\theta/n_0$,  and the heat capacity per particle at constant  density (constant volume $V_1$) and polarization,
\begin{equation}
c_{V_1}=T\left(\frac{\partial s_1}{\partial T}\right)_{\!\!Pn}=T\left(\frac{\partial s_1}{\partial\theta}\right)_{\!\!P}\left(\frac{\partial \theta}{\partial T}\right)_{\!\!n}=\theta\left(\frac{\partial s_1}{\partial\theta}\right)_{\!\!P}\, .
\label{eq:5.3S}
\end{equation}
With Eq.~\ref{eq:5.3S} we evaluate the last two terms in Eq.~\ref{eq:1.1s1} in terms of $c_{V_1}$ and solve for $\delta T/T_0=\delta\tilde{T}$,
\begin{equation}
\delta\tilde{T}=\frac{2}{3}\,\delta\tilde{n}-\frac{1}{c_{V_1}}\left(\frac{\partial s_1}{\partial P}\right)_{\!\!\theta}\delta P + \frac{\delta s_1}{c_{V_1}}.
\label{eq:1.7Sc}
\end{equation}
 
The thermodynamic relations obtained in this section determine all of the coefficients that appear in the spin hydrodynamic equations derived below. 

\subsection{Spin, Energy, and Heat Current Densities}
\label{sec:HeatSpin}

The central objective of the linear response model is to express the dynamics in terms of a minimal set of independent transport coefficients that can be extracted directly from experiment. To achieve this, the diffusive energy, heat, and spin current densities must be defined in a manner that is fully consistent with the thermodynamic relations of \S~\ref{sec:thermo} and with Onsager reciprocity. In this section, we establish these current densities and show that the hydrodynamic response can be written entirely in terms of the fundamental energy conductivity, energy-spin Seebeck coefficient, and spin diffusivity. Transport coefficients in different representations follow from these quantities through thermodynamic identities.  

\subsubsection{Diffusive Current Densities}

We begin by defining the diffusive spin current density ${\mathbf J}_i$  for each of the spin species, $i=\uparrow,\downarrow$. From Ref.~\cite{LandauFluids}, each density $n_i$ obeys
\begin{equation}
\dot{n}_i=-\nabla\cdot ({\mathbf v}_{\!s}\, n_i)-\nabla\cdot{\mathbf J}_i,
\label{eq:1.1S}
\end{equation}
with $\partial_t n_i\equiv\dot{n}_i$. Here, ${\mathbf v}_{\!s} \,n_i$ is the advective current density, arising from the stream velocity field ${\mathbf v}_{\!s}$. 

The total density $n=\sum_i n_i$ then obeys
\begin{equation}
\dot{n}=-\nabla\cdot\!({\mathbf v}_{\!s}\, n)-\nabla\cdot\sum_i{\mathbf J}_i.
\label{eq:2.4S}
\end{equation}

For a normal phase two-component spin-imbalanced Fermi gas, the velocity field  ${\mathbf v}_{\!s}({\mathbf{r}},t)$ for the total density $n=n({\mathbf{r}},t)$ is defined so that $n$ satisfies the continuity equation,
\begin{equation}
\dot{n} +\nabla\cdot(n {\mathbf v}_{\!s})=0,
\label{eq:continuity}
\end{equation}
which requires the last term in Eq.~\ref{eq:2.4S} to vanish for all ${\mathbf{r}}$. Hence, the sum of the diffusive spin current densities  ${\mathbf J}_{\uparrow\downarrow}$  vanishes,
\begin{equation}
\sum_i{\mathbf J}_i = {\mathbf J}_\uparrow+{\mathbf J}_\downarrow=0.
\label{eq:2.6S}
\end{equation}

Defining the magnetization $\sigma = n_\uparrow-n_\downarrow$ and using Eq.~\ref{eq:1.1S}, we see that
\begin{equation}
\dot{\sigma}=-\nabla\cdot({\mathbf v}_{\!s}\, \sigma)-\nabla\cdot{\mathbf J}_{\rm spin}.
\label{eq:2.11S}
\end{equation}
where the net diffusive spin current is defined by
\begin{equation}
{\mathbf J}_{\rm spin} = {\mathbf J}_\uparrow-{\mathbf J}_\downarrow.
\label{eq:2.6Sb}
\end{equation}
Now, the polarization $P=\sigma/n$, so that $\dot{P}=\dot{\sigma}/n-P\,\dot{n}/n$. Using $\nabla\cdot({\mathbf v}_{\!s}\, \sigma)=\nabla\cdot({\mathbf v}_{\!s}\, n P)$ with Eq.~\ref{eq:2.11S}~and the continuity equation Eq.~\ref{eq:continuity}, we find the evolution equation for the polarization,
\begin{equation}
(\partial_t+{\mathbf v}_{\!s}\!\cdot\!\nabla)P=-\frac{1}{n}\nabla\cdot{\mathbf J}_{\rm spin}.
\label{eq:3.5S}
\end{equation}

To determine the entropy production rate, we need to relate the heat current density to the diffusive energy and spin current densities. For this purpose, we consider the internal energy density ${\cal E}$, which obeys~\cite{LandauFluids}
\begin{equation}
\dot{\cal E}=-p\,\nabla\cdot{\mathbf v}_{\!s}-\nabla\cdot({\mathbf v}_{\!s}\,{\cal E})-\nabla\cdot{\mathbf J}_{\rm \epsilon}+\dot{q}_{\rm fric}.
\label{eq:1.3Se}
\end{equation}
Here, the first term is the power per unit volume arising from the work done by the gas, i.e., the pressure $p$ multiplied by the volume dilation rate.  ${\mathbf v}_{\!s}\,{\cal E}$  is the advective part of the internal energy current density and ${\mathbf J}_{\epsilon}$ is the diffusive energy current density. The last term $\dot{q}_{\rm fric}$ is the friction heating rate per unit volume arising from the stream flow.

Similarly, the entropy density $s=n s_1$, with $s_1$ the entropy per particle, obeys
\begin{equation}
\dot{s}=-\nabla\cdot({\mathbf v}_{\!s}\, s)+\frac{\dot{q}_{\rm heat}}{T},
\label{eq:1.2S}
\end{equation}
where ${\mathbf v}_{\!s}\, s$ is the advective entropy current density and $\dot{q}_{\rm heat}$ is the total heating rate per unit volume. We show below that $\dot{q}_{\rm heat}$ contains a conduction term  $-\nabla\cdot{\mathbf J}_{\rm heat}$ that is first order in small quantities, as well a second order terms, including $\dot{q}_{\rm fric}$, which contribute to the dissipation function.

Now we can relate the heat current  to the energy and spin currents using the fundamental thermodynamic relation for the internal energy density, 
\begin{equation}
d{\cal E}=T ds+\sum_i\mu_i\, dn_i\,,
\label{eq:1.4Sa}
\end{equation}
and the Gibbs-Duhem relation for the pressure, 
\begin{equation}
dp=s\,dT+\sum_i n_i\,d\mu_i\,.
\label{eq:1.4Sb}
\end{equation}
Together, Eqs.~\ref{eq:1.4Sa}~and~\ref{eq:1.4Sb} require
\begin{equation}
{\cal E}+p=T s+\sum_i\mu_i\,n_i\,.
\label{eq:1.5S}
\end{equation}
Defining $\dot{\cal E}\equiv\partial_t{\cal E}$, etc., in Eq.~\ref{eq:1.4Sa}, 
\begin{equation}
\dot{\cal E}=T\dot{s}+\sum_i\mu_i\,\dot{n}_i\, .
\label{eq:1.6S}
\end{equation}
We replace the time derivative terms in Eq.~\ref{eq:1.6S} with the right hand sides of Eqs.~\ref{eq:1.1S},~\ref{eq:1.3Se}~and~\ref{eq:1.2S}.  Expanding the advective contributions and regrouping yields
$$\dot{q}_{\rm fric}-\,\dot{q}_{\rm heat}-\nabla\cdot{\mathbf J}_{\rm \epsilon}+\sum_i\mu_i\nabla\cdot{\mathbf J}_i=\bigg({\cal E}+p-T s-\sum_i\mu_in_i\bigg)\,\nabla\!\cdot\!{\mathbf v}_{\!s}
+{\mathbf v}_{\!s}\!\cdot\!\bigg(\nabla{\cal E}-T\nabla s-\sum_i\mu_i\nabla n_i\bigg).$$
 Eqs.~\ref{eq:1.4Sa}~and~\ref{eq:1.5S} show that the sums in each of the brackets vanish. Hence,
\begin{equation}
\dot{q}_{\rm heat}=-\nabla\cdot{\mathbf J}_{\rm \epsilon}+\sum_i\mu_i\nabla\cdot{\mathbf J}_i+\dot{q}_{\rm fric}\, .
\label{eq:1.9S}
\end{equation}
Using Eqs.~\ref{eq:2.6S}~and~\ref{eq:2.6Sb},  ${\mathbf J}_{\uparrow,\downarrow}=\pm \,{\mathbf J}_{\rm spin}/2$. As $h=(\mu_\uparrow-\mu_\downarrow)/2$ from Eq.~\ref{eq:htheor}, we have
\begin{equation}
\sum_i\mu_i\nabla\cdot{\mathbf J}_i=h\,\nabla\cdot{\mathbf J}_{\rm spin}=\nabla\cdot (h\,{\mathbf J}_{\rm spin})-{\mathbf J}_{\rm spin}\cdot\nabla h\,.
\label{eq:summui}
\end{equation}
Then, Eq.~\ref{eq:1.9S} gives $\dot{q}_{\rm heat}=-\nabla\cdot{\mathbf J}_{\rm \epsilon}+h\,\nabla\cdot{\mathbf J}_{\rm spin} +\dot{q}_{\rm fric}$
or
\begin{equation}
\dot{q}_{\rm heat}=-\nabla\cdot{\mathbf J}_{\rm heat}- {\mathbf J}_{\rm spin}\cdot\nabla h +\dot{q}_{\rm fric},
\label{eq:2.1S}
\end{equation}
with the heat current density,
\begin{equation}
{\mathbf J}_{\rm heat}={\mathbf J}_{\epsilon}-h\,{\mathbf J}_{\rm spin}.
\label{eq:2.2S}
\end{equation}
 As we will expand the energy, heat and spin currents to first order in gradients of the relevant variables, ${\mathbf J}_{\rm heat}$  is first order in small quantities, while the last two terms in Eq.~\ref{eq:2.1S} are second order~\cite{LandauFluids} and contribute to the dissipation function. Note that ${\mathbf J}_{\epsilon}$ is denoted by ${\mathbf q}$ in Ref.~\cite{LandauFluids}.

Using Eq.~\ref{eq:2.1S} in Eq.~\ref{eq:1.2S} for  the entropy density  $s=ns_1$, we obtain
\begin{equation}
\dot{s}=-\nabla\cdot({\mathbf v}_{\!s}\,s)-\frac{\nabla\cdot{\mathbf J}_{\rm heat}}{T}-\frac{{\mathbf J}_{\rm spin}\cdot\nabla h}{T}+\frac{\dot{q}_{\rm fric}}{T}.
\label{eq:5.1S}
\end{equation}
Expanding the advective term in Eq.~\ref{eq:5.1S} and using the continuity equation~\ref{eq:continuity} for $n$, we have
\begin{equation}
nT(\partial_t+{\mathbf v}_{\!s}\!\cdot\!\nabla)\,s_1=-\nabla\cdot{\mathbf J}_{\rm heat}-{\mathbf J}_{\rm spin}\cdot\nabla h+\dot{q}_{\rm fric}\,.
\label{eq:2.9S}
\end{equation}

\subsubsection{Expansion of the Current Densities}
\label{sec:natvar}

We extract from the data the transport coefficients for the fundamental energy and spin current densities, ${\mathbf J}_{\epsilon}$ and ${\mathbf J}_{\rm spin}$. Following  Ref.~\cite{LandauFluids}, we find the thermodynamic forces that determine the current densities from the total entropy production rate $\dot{S}=\int d^3{\mathbf r}\,\dot{s}$. Using Eq.~\ref{eq:5.1S} with $\nabla\cdot{\mathbf J}_{\rm heat}/T=\nabla\cdot({\mathbf J}_{\rm heat}/T)-{\mathbf J}_{\rm heat}\cdot\nabla(1/T)$,
\begin{equation}
\dot{S}=-\int d^3{\mathbf r}\,\left[\nabla\left(\frac{-1}{T}\right)\cdot{\mathbf J}_{\rm heat}+\frac{\nabla h}{T}\cdot{\mathbf J}_{\rm spin}\right]=-\int d^3{\mathbf r}\,\left[\nabla\left(\frac{-1}{T}\right)\cdot{\mathbf J}_{\epsilon}+\nabla\left(\frac{h}{T}\right)\cdot{\mathbf J}_{\rm spin}\right]\,,
\label{eq:thermoforces}
\end{equation} 
where we drop $\dot{q}_{\rm fric}$ for brevity. 

The coefficients of the currents in Eq.~\ref{eq:thermoforces} determine the proper thermodynamic forces for expansion of the current densities. For the energy and spin current densities, we expand in gradients of the variables $-1/T$ and $h/T$,
\begin{eqnarray}
{\mathbf J}_{\epsilon}&=&-\kappa_\epsilon\,T^2\nabla(-1/ T)\,-\,P_\epsilon\,T\nabla (h/T)\nonumber\\
{\mathbf J}_{\rm spin}&=&-S_\epsilon\,T^2\nabla(-1/T)\,-\,2\sigma_\epsilon\,T\nabla (h/T)\,.
\label{eq:currents1}
\end{eqnarray}
Here, we have chosen the extra factors of $T$ so that $\kappa_\epsilon$ is the energy conductivity at constant $h/T$ and $\sigma_\epsilon$ is a spin conductivity. As in Ref.~\cite{LandauFluids}, the Onsager relation requires the off-diagonal coefficients to be symmetric,   $P_\epsilon\,T = S_\epsilon\,T^2$, reducing the number of fit parameters. We define $S_\epsilon$ as the energy-spin Seebeck coefficient, with $P_\epsilon=T S_\epsilon$ the corresponding energy-spin Peltier coefficient. We find that this choice of variables and transport coefficients leads to important constraints in the data analysis, as discussed in \S~\ref{sec:Seebeck}.  

Using Eq.~\ref{eq:htheor}, we can expand $T\,\nabla (h/T)$ as
\begin{equation}
T\,\nabla\left(\frac{h}{T}\right)=\left(\frac{\partial h}{\partial P}\right)_{nT}\left[\nabla P + c_{sn}\left(\frac{\nabla n}{n}-\frac{3}{2}\,\frac{\nabla T}{T}\right)\right]\,,
\label{eq:TnablahoverT}
\end{equation}
where $c_{sn}=n(\partial h/\partial n)_{PT}/(\partial h/\partial P)_{nT}$ is given by
\begin{equation}
c_{sn}\equiv\frac{2}{3}\,\frac{f_h-\theta(\partial f_h/\partial\theta)_P}{(\partial f_h/\partial P)_\theta}\,.
\label{eq:csnS}
\end{equation}
As discussed in \S~\ref{sec:expt}, the thermodynamic quantitiy $c_{sn}=-n(\partial P/\partial n)_{hT}$ is measured in the experiments.
In the high temperature limit, where $f_h=(\theta/2)\ln[(1+P)/(1-P)]$, we see that $c_{sn}\rightarrow 0$ and $T\,\nabla\left(\frac{h}{T}\right)\rightarrow\left(\frac{\partial h}{\partial P}\right)_{nT}\nabla P$.

For the spin hydrodynamics model, we need to expand the heat and spin current densities in gradients of the density, polarization, and temperature.  We begin by making contact with the conventional description of spin caloritronic transport for the heat and spin current densities.  Using Eq.~\ref{eq:thermoforces}, we expand the current densities in terms of $T^2\nabla (-1/T)=\nabla T$ and $T\nabla h/T=\nabla h$. Using notation similar to that of Ref.~\cite{KimHuseSpinDiff}, 
\begin{eqnarray}
{\mathbf J}_{\rm heat}&=&{\mathbf J}_{\epsilon}-h\,{\mathbf J}_{\rm spin}=-\kappa_T\nabla T\,-\,P_s\,\nabla h\nonumber\\
{\mathbf J}_{\rm spin}&=&-S_s\,\nabla T\,-\,2\sigma_s\nabla h\,,
\label{eq:currents2}
\end{eqnarray}
which defines the spin conductivity $\sigma_s$ and the heat-spin Seebeck coefficient $S_s$.
From the heat current density, we identify $\kappa_T$ as the thermal conductivity and $P_s$ as the heat-spin Peltier coefficient.
From the spin current density for $\nabla T=0$, we have ${\mathbf J}_{\rm spin}=-2\sigma_s\nabla h=-\sigma_s\nabla(\mu_\uparrow-\mu_\downarrow)$, which  defines $\sigma_s$. 
 With ${\mathbf J}_{\rm spin}=-D_s\nabla(n_\uparrow-n_\downarrow)$  for $\nabla T=0$ and $\nabla n=0$, the spin conductivity $\sigma_s$ is related to the spin diffusivity $D_s$ by
\begin{equation}
n D_s=2\sigma_s \left(\frac{\partial h}{\partial P}\right)_{Tn}\,.
\label{eq:DsS}
\end{equation}

Comparing the coefficients of $\nabla T$ and $\nabla h$ for the spin currents in Eq.~\ref{eq:currents2} and Eq.~\ref{eq:currents1}, we have $\sigma_\epsilon=\sigma_s$. For the heat-spin Seebeck coefficient, we obtain
\begin{equation}
S_s=S_\epsilon-2\sigma_s\frac{h}{T}\,,
\label{eq:Ss}
\end{equation}
and for the heat-spin Peltier coefficient,
\begin{equation}
P_s=P_\epsilon-2\sigma_s h=T S_s\,.
\end{equation}
We see that the Onsager relation $P_\epsilon = T S_\epsilon$ for the variables $-1/T$ and $h/T$ is consistent with the Onsager relation $P_s=TS_s$ for the $h-T$ variables, as expected from Eq.~\ref{eq:thermoforces}. The thermal conductivity is given by
\begin{equation}
\kappa_T=\kappa_\epsilon-2hS_\epsilon+2\sigma_s\frac{h^2}{T}\,.
\label{eq:kappaT}
\end{equation}

Now we are ready to express the heat and spin current densities in terms of the gradients of $n$, $P$, and $T$ and the fundamental transport properties $\kappa_\epsilon$, $D_s$, and $S_\epsilon$. This is accomplished by rewriting the heat and spin current densities of Eq.~\ref{eq:currents2} in a form similar to the $P-T$ representation of Ref.~\cite{KimHuseSpinDiff}. Using ${\mathbf J}_{\rm heat}={\mathbf J}_{\epsilon}-h\,{\mathbf J}_{\rm spin}$ from Eq.~\ref{eq:2.2S},  we expand the energy and spin current densities of Eq.~\ref{eq:currents1} using Eq.~\ref{eq:TnablahoverT}. Including the essential $c_{sn}$ terms, which vanish in the high temperature limit of Ref.~\cite{KimHuseSpinDiff}, we obtain the general form
\begin{eqnarray}
{\mathbf J}_{\rm heat}&=&-\kappa_T'\nabla T\,-\,n P_s'\,\left(\nabla P + c_{sn}\frac{\nabla n}{n}\right)\nonumber\\
{\mathbf J}_{\rm spin}&=&-S_s'\,\nabla T\,-\,n D_s\,\left(\nabla P + c_{sn}\frac{\nabla n}{n}\right) \,.
\label{eq:currents3}
\end{eqnarray}

We write the transport coefficients in Eq.~\ref{eq:currents3} in terms of dimensionless transport coefficients, denoted by a tilde, as follows,
\begin{eqnarray}
&\kappa_T'=\tilde{\kappa}'_T\,\frac{k_B}{m}\,\hbar n\hspace{0.25in}P_s'=\tilde{P}_s\left(\frac{\partial h}{\partial P}\right)_{\!\!nT}\frac{\hbar}{m}\nonumber\\
&S_s'=\tilde{S}_s'\,\frac{n}{T}\frac{\hbar}{m}\hspace{0.25in}D_s\equiv\tilde{D}_s\,\frac{\hbar}{m}
\hspace{0.25in}\kappa_\epsilon=\tilde{\kappa}_\epsilon\,\frac{k_B}{m}\,\hbar n\hspace{0.25in}S_\epsilon=\tilde{S}_\epsilon\,\frac{n}{T}\frac{\hbar}{m}\,.
&\nonumber\\
\label{eq:Transp}
\end{eqnarray}

The dimensionless Seebeck coefficient $\tilde{S}_s'$ in Eq.~\ref{eq:Transp} is 
\begin{equation}
\tilde{S}_s'=\tilde{S}_\epsilon-\tilde{D}_s\,\frac{3}{2}\,c_{sn}\,.
\label{eq:4.9Ssprime}
\end{equation}

For the thermal conductivity, we find
\begin{equation}
\tilde{\kappa}_T'=\tilde{\kappa}_\epsilon+\tilde{S}_\epsilon\left[\left(\frac{\partial f_h}{\partial\theta}\right)_{\!P}\!\!-2\,\frac{f_h}{\theta}\right]+\tilde{D}_s\frac{f_h}{\theta}\frac{3}{2}\,c_{sn}\,.
\label{eq:4.7kappaTprime}
\end{equation}

The dimensionless Peltier coefficient in Eq.~\ref{eq:Transp} is
\begin{equation}
\tilde{P}_s=\tilde{S}_\epsilon-\tilde{D}_s\,\frac{f_h}{(\partial f_h/\partial P)_\theta}\,.
\label{eq:4.9Ps}
\end{equation}

With the definitions $S_s=\tilde{S}_s\,\frac{n}{T}\frac{\hbar}{m}$ and $P_s=\tilde{P}_s\,n\frac{\hbar}{m}$, we see that the heat-spin Seebeck coefficient of Eq.~\ref{eq:Ss} corresponds to
\begin{equation}
\tilde{S}_s=\tilde{S}_\epsilon-\tilde{D}_s\,\frac{f_h}{(\partial f_h/\partial P)_\theta}
\label{eq:tildeSs}
\end{equation}
and that $\tilde{P}_s=\tilde{S}_s$, which is just the Onsager relation $P_s=T S_s$ in the $h-T$ representation, as discussed above.

We now have  all of the transport coefficients that appear in the heat and spin currents defined in terms of the thermodynamic function $f_h(P,\theta)$ and the fundamental transport properties $\tilde{\kappa}_\epsilon$, $\tilde{S}_\epsilon$, and $\tilde{D}_s$, which are the central fit parameters of the model.  Eq.~\ref{eq:currents3} determines the heat and spin currents in the form that will be used to derive the linear response evolution equations.
 
\subsection{Transport Properties from Boltzmann Theory}
\label{sec:Boltzmann}

The hydrodynamic equations derived in \S~\ref{sec:Hydro} are formulated in terms of a small set of transport coefficients and are independent of any particular microscopic theory. To provide a quantitative point of comparison with the experimental results, we evaluate these transport coefficients within a microscopic kinetic theory. This approach yields parameter-free predictions for the spin diffusivity and the energy-spin Seebeck coefficient, allowing direct comparison with the values extracted from the hydrodynamic analysis.

The temperature and polarization dependence of the transport coefficients is calculated using a quantum Boltzmann equation that incorporates medium-enhanced scattering through the many-body T-matrix. Although the quantum critical regime of the unitary Fermi gas is not characterized by long-lived quasiparticles, the Boltzmann equation nevertheless emerges as the leading-order kinetic description in a controlled expansion in the inverse number of fermionic flavors, $1/N$~\cite{Enss2012}. Within this framework, medium scattering substantially enhances the scattering rate compared with vacuum two-body collisions near quantum degeneracy and has been shown to significantly improve the description of spin transport~\cite{Enss2012}. The present calculation follows the same large-$N$ formalism, generalized here to spin-imbalanced gases.

We therefore consider the quantum Boltzmann equation
\begin{align}
    S_\sigma(\mathbf p)=I^\sigma_{\rm coll}(\mathbf p),
\end{align}
where $\sigma=\pm 1$ denotes the spin states $\uparrow,\downarrow$. The streaming terms are generated by gradients in temperature and relative chemical potential,
\begin{align}
    S_\sigma(\mathbf{p})=-n_F^\sigma(\mathbf{p})(1-n_F^\sigma(\mathbf{p}))\mathbf{v}_p\left[\left(\epsilon_p-w\right)\frac{1}{k_B}\nabla \left(-\frac{1}{T}\right)-\left(\sigma-P\right)\frac{1}{k_B}\nabla \frac{h}{T}\right]\,.
\end{align}
Here, 
\begin{align}
w=\mu+T\tilde{s}+P h
\end{align}
denotes the enthalpy per particle. Linearizing the distribution function around local equilibrium according to
\begin{align}
n_\sigma(\mathbf x,t,\mathbf p)
=
n^\sigma_{F_{\rm loc}}(\mathbf x,t,\mathbf p)
+
n_F^\sigma(p)
\left(1-n_F^\sigma(p)\right)
\phi_\sigma(\mathbf p),
\end{align}
with
\begin{align}
    n_{F_\text{loc}}^\sigma(\mathbf{x},t,\mathbf{p})=\left[\exp\left((\epsilon_p-\mu_\sigma(\mathbf{x}))/(k_BT(x))\right)+1\right]^{-1}\,,
\end{align}
and expressing the scattering cross section in terms of the many-body $T$-matrix, the linearized collision integral takes the form
\begin{align}
\begin{split}
    I_\text{coll}^\sigma=&\int\frac{d p'^3}{(2\pi\hbar)^3}|\check{\mathbf{v}}_p-\check{\mathbf{v}}_{p'}|\int d \Omega_{k'}\left|\mathcal{T}(\mathbf{p}+\mathbf{p}',\check{\epsilon}_\mathbf{p}^\sigma+\check{\epsilon}_{\mathbf{p'}}^{\bar\sigma})\right|^2
    n_\uparrow(\mathbf{p}+\mathbf{p}'-\mathbf{q})\,n_\downarrow(\mathbf{q})\\&\hspace{0.5in}\times(1-n_{\bar\sigma}(\mathbf{p}'))(1-n_\sigma(\mathbf{p}))
    \left[\phi_\sigma(\mathbf{p})+\phi_{\bar\sigma}(\mathbf{p}')-\phi_\downarrow(\mathbf{q})-\phi_\uparrow(\mathbf{q'})\right]\,.
\end{split}
\end{align}
Here $\check{\epsilon}_p$ and $\check{\mathbf v}_p$ denote the quasiparticle energy and group velocity. To leading order in a systematic expansion in the inverse number of fermionic flavors $1/N$, these are approximated by their free-gas expressions, while medium effects are incorporated through the scattering amplitude.

To leading order in $1/N$, the $T$-matrix of the imbalanced Fermi gas is given by
\begin{eqnarray}
\mathcal{T}(\hat{Q},\hat{\omega})&=&\left[\frac{1}{a}-\frac{\sqrt{m_0 k_B T}}{\hbar}\sqrt{\hat{Q}^2/2-\hat{\omega}-i0^+-2\hat{\mu}}\right.\\
& &\hspace{-0.25in}\left.+\frac{\sqrt{2m_0k_BT}}{\pi\hbar}\int_0^\infty\!\! d\hat{p}\,\frac{\hat{p}}{\hat{Q}}\left(n_F^\uparrow(\hat{p}^2)+n_F^\downarrow(\hat{p}^2)\right)
    \ln\left(\frac{\hat{\omega}+i0^+-\hat{p}^2-(\hat{p}-\hat{Q})^2+2\hat{\mu}}{\hat{\omega}+i0^+-\hat{p}^2-(\hat{p}+\hat{Q})^2+2\hat{\mu}}\right)\right]^{-1}\,,\nonumber
\end{eqnarray}
where the last term describes medium scattering.

Without loss of generality, we take the applied gradients to point along the $z$-direction. The linearized Boltzmann equation can then be solved variationally by expanding $\phi_\sigma$ in an orthogonal basis. Restricting ourselves to the lowest velocity moment, which is equivalent to the moments method employed in Ref.~\cite{KimHuseSpinDiff}, we write~\cite{Enss2012,Frank2019}
\begin{eqnarray}
    \phi_\sigma(\mathbf{p})&=&{\rm v}_z\left(A\left(\epsilon_p-w\right)-B\left(\sigma-P\right)+C+ D \left(\sigma \epsilon_p-\frac{5}{3}\frac{{\cal E}_\uparrow-{\cal E}_\downarrow}{n}\right)\right)\nonumber\\
    &\equiv& A\phi_A+B\phi_B+C\phi_C + D \phi_D\,.
\end{eqnarray}
It is convenient to work in the flow frame defined by a vanishing particle current, which fixes $C=0$ while leaving the remaining coefficients unconstrained. The linearized Boltzmann equation is then reduced to a finite-dimensional linear algebra problem by projecting onto $\phi_{A,B,D}$. The transport coefficients follow directly from expressing the heat and spin currents in terms of the resulting coefficients $A$, $B$, and $D$.

We emphasize that in the quantum critical regime, the density of the unitary Fermi gas differs substantially from that of a free gas, implying sizable changes in transport coefficients expressed in density units. To ensure thermodynamic consistency, all thermodynamic quantities entering the calculation are therefore evaluated within the Luttinger--Ward framework discussed above.

No such issue arises at high temperatures, where medium corrections become negligible, and the present formalism reduces to the moments expansion of Ref.~\cite{KimHuseSpinDiff}, from which the transport coefficients can be obtained analytically, as we discuss next in \S~\ref{sec:highT}.

\subsubsection{Transport Properties in the High Temperature Kinetic Theory Limit}
\label{sec:highT}

In the high temperature limit, $h=\frac{k_BT}{2}\,\ln [(1+P)/(1-P)]$.  Then $T\,\nabla\left(\frac{h}{T}\right)\rightarrow\left(\frac{\partial h}{\partial P}\right)_{nT}\nabla P$ in the current densities of Eq.~\ref{eq:currents1}. In this case, the energy and spin current densities can be evaluated analytically using a linearized two-body Boltzmann equation~\cite{KimHuseSpinDiff}, where the coefficients of $\nabla T$ and $\nabla P$ determine $D_s$, $S_\epsilon$, $P_\epsilon$, and $\kappa_\epsilon$. 

We write the hydrodynamic and spin transport properties of the unitary Fermi gas in terms of the spin diffusivity obtained by a variational method in Ref.~\cite{BruunSpinDiff}, 
\begin{equation}
D_{s0}=\frac{9\pi^{3/2}}{32\sqrt{2}}\,\left(\frac{T}{T_F}\right)^{3/2}\!\frac{\hbar}{m}\,.
\label{eq:Ds0}
\end{equation}

Using the moments method of Ref.~\cite{KimHuseSpinDiff} and Eq.~\ref{eq:currents1} in the high temperature limit, we find the spin diffusivity
\begin{equation} 
D_s=\frac{39}{38}\,D_{s0}\,,
\label{eq:DshighT}
\end{equation}
which is independent of $P$. We obtain the energy-spin Seebeck coefficient, 
\begin{equation} 
S_\epsilon=\frac{5}{19}\,P\,\frac{n}{T}\,D_{s0}
\label{eq:DshighT}
\end{equation}
and the energy-spin Peltier coefficient $P_\epsilon$, which obeys the Onsager relation $P_\epsilon=T S_\epsilon$ as it should.

Using the moments method, we also determine the shear viscosity,
\begin{equation}
\eta =\eta_0\,\frac{1+\frac{2}{5}P^2}{1-P^2}\,,
\label{eq:shearP}
\end{equation}
where $\eta_0/(nm) = 5/2\,D_{s0}$~\cite{BruunViscousNormalDamping,BluhmSchaeferModIndep,BluhmSchaeferLocalViscosity}. 
In terms of $\eta_0$, the energy conductivity is 
\begin{equation}
\kappa_\epsilon=\frac{15}{4}\frac{k_B}{m}\,\eta_0\,\frac{1+\frac{16}{57}P^2}{1-P^2}\,.
\label{eq:kappaeps}
\end{equation}
For $P=0$, Eq.~\ref{eq:kappaeps} is in agreement with Ref.~\cite{BrabySchaeferThermalCond} as it should be.

The first sound diffusivity $D_1$ is discussed in \S~\ref{sec:D1}. There,  we use Eqs.~\ref{eq:shearP}~and~\ref{eq:kappaeps}
to obtain the polarization dependence of $D_1$ in the kinetic theory limit.

These microscopic predictions provide parameter-free benchmarks for comparison with the transport coefficients extracted from the experimental data.

\subsection{Hydrodynamic Linear Response for a Spin-Imbalanced Normal Fluid}
\label{sec:Hydro}

We now combine the thermodynamic relations of \S~\ref{sec:thermo} with the transport currents established in \S~\ref{sec:HeatSpin} to derive the coupled linear-response equations governing the  density, polarization, and temperature perturbations. The resulting evolution equations provide the theoretical model used throughout the main text to analyze the data. A key consequence of this formulation is that the dynamics depend only on the fundamental transport coefficients introduced in \S~\ref{sec:HeatSpin}, leading to experimentally useful constraints that enable a robust extraction of the energy-spin Seebeck coefficient, spin diffusivity, and first-sound diffusivity. 
The central result of this section is Eq.~\ref{eq:2.1Modes} of \S~\ref{sec:HydroA}, which gives the final evolution equations in terms of the energy conductivity, $\kappa_\epsilon$, energy-spin Seebeck coefficient $S_\epsilon$, and spin diffusivity $D_s$ of \S~\ref{sec:natvar}. As shown in \S~\ref{sec:Seebeck}, an important constraint on $S_\epsilon$ emerges from this unique choice of fundamental transport coefficients, enabling robust determination of the transport properties.

\subsubsection{Evolution Equations}
\label{sec:HydroA}

For a normal phase two-component spin-imbalanced Fermi gas, we define the velocity field  ${\mathbf v}_{\!s}({\mathbf{r}},t)$ for the total density $n=n({\mathbf{r}},t)$, so that $n$ satisfies the continuity equation Eq.~\ref{eq:continuity}, 
\begin{equation}
\partial_t n +\partial_i(n {\rm v}_{si})=0,
\label{eq:1.3S}
\end{equation}
where a sum over $i=x,y,z$ is implied.

The mass flux (momentum density) is $\rho {\rm v}_{si}$, with $\rho = mn$ the mass density. The momentum density and corresponding momentum flux $\rho {\rm v}_{si} {\rm v}_{s j}$ obey
\begin{equation}
\partial_t (\rho\,{\rm v}_{si}) +\partial_j (\rho\,{\rm v}_{si} {\rm v}_{sj})=-\partial_i p -n\,\partial_i U+\partial_j\sigma'_{ij},
\label{eq:1.5Sn}
\end{equation}
Here, $-\partial_i p-n\,\partial_i U$ is the force per unit volume arising from the pressure $p$ and the externally applied potential $U({\mathbf{r}},t)$. The last term describes the dissipative forces, which arise generally from the shear viscosity  $\eta$ and the bulk viscosity $\xi_B$, with $\sigma'_{ij}=\eta\,\sigma_{ij}+\xi_B\,\delta_{ij}\nabla\cdot{\mathbf{v}_s}$ and $\sigma_{ij}\equiv\partial_i{\rm v}_{sj}+\partial_j{\rm v}_{si}-2\,\delta_{ij}\nabla\cdot{\mathbf{v}_s}/3$.  We  have shown experimentally that for a unitary Fermi gas, $\xi_B$ is negligible compared to $\eta$~\cite{ElliottScaleInv}, consistent with predictions that $\xi_B=0$ for a unitary gas~\cite{SonBulkViscosity,StringariBulk}. Taking the divergence of Eq.~\ref{eq:1.5Sn}, and using Eq.~\ref{eq:1.3S}, we immediately obtain
\begin{equation}
-\partial_t^2\rho+\partial_i\partial_j (\rho\,{\rm v}_{si} {\rm v}_{s j})=-\partial_i^2 p-\partial_i(n\,\partial_iU)+\partial_i\partial_j\sigma'_{ij}.
\label{eq:2.7S}
\end{equation}

The applied potential in Eq.~\ref{eq:2.7S} takes the form $U({\mathbf{r}},t)=U_0({\mathbf{r}})+\delta U({\mathbf{r}},t)$, where $U_0$ is the box potential, which creates a uniform density sample. We are interested in the  hydrodynamic linear response to the perturbing external potential $\delta U({\mathbf{r}},t)$, which leads to first order changes in the density $n=n_0+\delta n({\mathbf{r}},t)$ and pressure $p=p_0+\delta p({\mathbf{r}},t)$.
Here, $n_0({\mathbf{r}})$ and $p_0({\mathbf{r}})$ are the equilibrium (time independent) density and pressure arising from confinement in the box trap potential, $U_0({\mathbf{r}})$. In equilibrium, the velocity field ${\mathbf v}_{s}({\mathbf{r}},t)=0$ and Eq.~\ref{eq:1.5Sn} requires balance of the forces per unit volume arising from the box trap and the pressure,
\begin{equation}
-\partial_i p_0({\mathbf{r}})-n_0({\mathbf{r}})\,\partial_i U_0({\mathbf{r}})=0.
\label{eq:1.3b}
\end{equation}
Substituting $U=U_0+\delta U$  into  Eq.~\ref{eq:2.7S} and retaining terms to first order in small quantities, we obtain
\begin{equation}
\partial_t^2\delta n=\frac{1}{m}\nabla^2\,\delta p+\frac{1}{m}\nabla\cdot[n_0({\mathbf{r}})\,\nabla\delta U+\delta n\,\nabla U_0]-\frac{1}{m}\partial_i\partial_j\sigma'_{ij}.
\label{eq:3.7S}
\end{equation}
Here, the second term on the left side of Eq.~\ref{eq:2.7S} is negligible, as the velocity field is first order in small quantities.

To evaluate the last term in Eq.~\ref{eq:3.7S}, we assume that the dissipative forces are small compared to the conservative forces and that the density $n_0$ slowly varies in the region of interest. Then we can ignore the spatial derivatives of $\eta$, $\xi_B$, and $n_0$, yielding
$\partial_i\partial_j\sigma'_{ij}\simeq (4/3\,\eta+\xi_B)\,\nabla^2(\nabla\cdot{\mathbf{v}}_s)$, where  $\xi_B=0$  for a unitary Fermi gas as noted above.
The velocity field is eliminated using $\nabla\cdot{\mathbf{v}}_s\simeq -\partial_t\delta n/n_0=-\delta\dot{n}/n_0$, which follows from  Eq.~\ref{eq:1.3S}.

For our experiments, the perturbing potential $\delta U$ is used to create the initial spatially periodic density perturbation and is extinguished during the evolution. Further, the evolution of the primary Fourier component is measured near the center of the density profile for time scales that avoid reflections from the edges of the box potential $U_0$. Hence, we omit the $U_0$ and $\delta U$  terms in Eq.~\ref{eq:3.7S}. Using Eq.~\ref{eq:deltap} for $\delta p$ and dividing by $n_0$, we obtain
\begin{equation}
\delta\ddot{\tilde{n}}=c_T^2\,\nabla^2(\,\delta\tilde{n}\,+\,3/2\, \epsilon_{LP}\,\delta\tilde{T}\,)+ \frac{{\rm v}_F^2}{2}\!\left(\frac{\partial f_p}{\partial P}\right)_{\!\!\theta}\nabla^2\delta P
+\frac{4}{3}\,\tilde{\eta}\frac{\hbar}{m}\,\nabla^2\delta\dot{\tilde{n}}\,.
\label{eq:2.5S}
\end{equation}
Here, we have used the dimensionless variables $\delta\tilde{n}=\delta n/n_0$ and $\delta\tilde{T}\equiv\delta{T}/T_0$ from Eq.~\ref{eq:1.2Sb}. The viscosity is $\eta=\tilde{\eta}\,\hbar\/n_0$, where $\tilde{\eta}$ a dimensionless fit parameter.

The polarization change $\delta P$ evolves according to Eq.~\ref{eq:3.5S}. To first order in small quantities,
\begin{equation}
\delta\dot{P}=-\frac{1}{n_0}\nabla\cdot{\mathbf J}_{\rm spin},
\label{eq:3.5aS}
\end{equation}
where ${\mathbf J}_{\rm spin}$ is the diffusive part of the spin current, discussed below.

The evolution of $\delta\tilde{T}$ is found by differentiating Eq.~\ref{eq:1.7Sc} of \S~\ref{sec:thermo},
\begin{equation}
\delta\dot{\tilde{T}}=\frac{2}{3}\,\delta\dot{\tilde{n}}-\frac{1}{c_{V_1}}\left(\frac{\partial s_1}{\partial P}\right)_{\!\!\theta}\delta\dot{P} + \frac{\dot{\delta s_1}}{c_{V_1}}.
\label{eq:1.9Sa}
\end{equation}

The heating rate per particle, $T\delta\dot{s}_1$ is found from Eq.~\ref{eq:2.9S}. To first order in small quantities,
\begin{equation}
\frac{\delta\dot{s}_1}{c_{V_1}}=-\frac{1}{n_0\,T_0\,c_{V_1}}\nabla\cdot{\mathbf J}_{\rm heat}.
\label{eq:4.2ST}
\end{equation}
Note that the viscous heating rate is second order in the stream velocity components ${\rm v}_{si}$, which is negligible compared to the heating rate arising from the heat current density
${\mathbf J}_{\rm heat}$, as discussed  in \S~\ref{sec:HeatSpin}.

To extract the spin and heat transport properties from our measurements, we use  Eq.~\ref{eq:currents3} from \S~\ref{sec:natvar}
to express heat and spin current densities to first order in the gradients of $\delta T$ and the measured variables $\delta P$, and $\delta\tilde{n}$,   
\begin{equation}
{\mathbf J}_{\rm heat}=-\kappa'_T\,\nabla\delta T-n_0 P'_s\, (\nabla\delta P+c_{sn}\nabla\delta\tilde{n})
\label{eq:1.10heat}
\end{equation}
\begin{equation}
{\mathbf J}_{\rm spin}=-S'_s\,\nabla\delta T-n_0 D_s\, (\nabla\delta P+c_{sn}\nabla\delta\tilde{n})\,.
\label{eq:1.10spin}
\end{equation}
In this representation,  $c_{sn}\equiv -\delta P(q,0)/\delta\tilde{n}(q,0)$  assures that the spin and heat currents vanish in thermal equilibrium at $t=0$ as they should.  As discussed in \S~\ref{sec:expt} and shown in Fig.~\ref{fig:CsnvsP},  $c_{sn}$   is a measured thermodynamic quantity given by Eq.~\ref{eq:1.15}, which can be compared to predictions using Eq.~\ref{eq:csnS} and the theoretical equation of state for $h$, Eq.~\ref{eq:htheor}.

Using Eq.~\ref{eq:Transp}, we have 
\begin{equation}
S_s'=\tilde{S}'_s\, \frac{n_0}{T_0}\,\frac{\hbar}{m}\,.
\label{eq:3.2heatspin}
\end{equation}
where the dimensionless spin Seebeck coefficient $\tilde{S}'_s$ given by Eq.~\ref{eq:4.9Ssprime} in \S~\ref{sec:natvar}. 

Writing the spin diffusion constant as $D_s=\tilde{D}_s\hbar/m$, and using Eq.~\ref{eq:3.2heatspin}, we determine the spin current density, Eq.~\ref{eq:1.10spin}. Using this in Eq.~\ref{eq:3.5aS}, we find
\begin{equation}
\delta\dot{P}=\tilde{S}'_s\,\frac{\hbar}{m}\,\nabla^2\delta\tilde{T}+\tilde{D}_s\,\frac{\hbar}{m}\,\nabla^2(\delta P+c_{sn}\,\delta\tilde{n})\,,
\label{eq:3.12heatspin}
\end{equation}
where $\tilde{S}_s'=\tilde{S}_\epsilon-\tilde{D}_s\,\frac{3}{2}\,c_{sn}$ from Eq.~\ref{eq:4.9Ssprime}.
As the densities are initially static, we must have $\delta\dot{P}=0$ at $t=0$. We see that this is enforced by Eq.~\ref{eq:3.12heatspin}, since at $t=0$,  the last term vanishes, as noted above, and $\delta\tilde{T}=0$. 

For the heat current of Eq.~\ref{eq:1.10heat},  the transport coefficients $\kappa_T'$ and $P_s'$ in are given by Eq.~\ref{eq:Transp} in \S~\ref{sec:natvar}.
Using these results, we evaluate Eq.~\ref{eq:4.2ST}, which gives the last term of Eq.~\ref{eq:1.9Sa}, yielding
\begin{equation}
\delta\dot{\tilde{T}}=\frac{2}{3}\,\delta\dot{\tilde{n}}+C_{T\!P}\,\delta\dot{P} +\frac{\tilde{\kappa}'_T}{\tilde{c}_{V_1}}\frac{\hbar}{m}\,\nabla^2\delta\tilde{T}+
\tilde{P}_s\,C_{\!P_s}\,\frac{\hbar}{m}\,(\nabla^2\delta P+c_{sn}\nabla^2\delta\tilde{n})\,.
\label{eq:4.7heatspin}
\end{equation}
Eq.~\ref{eq:4.7heatspin} below shows that $\delta\dot{\hat{T}}=0$ at $t=0$.
Here, $\tilde{\kappa}'_T$ is given by Eq.~\ref{eq:4.7kappaTprime} and $\tilde{P}_s$ is given by Eq.~\ref{eq:4.9Ps}. We have defined
\begin{equation}
C_{T\!P}\equiv -\frac{1}{\tilde{c}_{V_1}}\left(\frac{\partial f_{s_1}}{\partial P}\right)_{\!\!\theta}>0
\label{eq:4.2heatspin}
\end{equation}
and
\begin{equation}
C_{\!P_s}\equiv \frac{1}{\tilde{c}_{V_1}}\frac{1}{\theta}\!\left(\frac{\partial f_h}{\partial P}\right)_{\!\!\theta}\,,
\label{eq:4.6heatspin}
\end{equation}
where   $\tilde{c}_{V_1}$ is the  heat capacity per particle at constant volume and $f_{s_1}$ is the entropy per particle, in units of $k_B$. 

Eqs.~\ref{eq:2.5S},~\ref{eq:3.12heatspin}~and~\ref{eq:4.7heatspin} describe the evolution of the perturbed spin-imbalanced unitary Fermi gas. We employ a perturbation that varies along one axis $z$, where the variables are $\delta\tilde{n}(z,t)$, $\delta\tilde{T}(z,t)$, and $\delta P(z,t)$, so that $\nabla^2\rightarrow\partial_z^2$.

As the perturbation is periodic with a wavevector $q$, we employ a spatial Fourier transform of these equations to model the data, which formally replaces $\nabla^2$ with $-q^2$,  yielding the corresponding time-dependent evolution equations for the variables $\delta\tilde{n}(q,t)$, $\delta\tilde{T}(q,t)$, and $\delta P(q,t)$. In these equations, it is convenient to define the isothermal sound frequency $\omega_T=  c_T q$, the Fermi frequency, $\omega_F={\rm v}_F q$ and the natural diffusion rate $\gamma_0=\hbar q^2/m$. 

Now we can write the evolution equations for the Fourier components that form the theoretical model used throughout the main text. Using $\delta\tilde{n}\equiv\delta\tilde{n}(q,t)$,  $\delta\tilde{T}\equiv\delta\tilde{T}(q,t)$ and $\delta P\equiv\delta P(q,t)$, we find

\begin{eqnarray}
\delta\ddot{\tilde{n}}&=&-\omega_T^2(\delta\tilde{n}+3/2\,\epsilon_{LP}\,\delta\tilde{T})-\Omega_{np}^2\,\delta P-\gamma_\eta\,\delta\dot{\tilde{n}}\nonumber\\
& &\nonumber\\
\delta\dot{\tilde{T}}&=&\frac{2}{3}\,\delta\dot{\tilde{n}}+C_{T\!P}\,\delta\dot{P}-\gamma_\kappa\,\delta\tilde{T}-\Gamma_{\!P_s}\,(\delta P+c_{sn} \delta\tilde{n})\nonumber\\
& &\nonumber\\
\delta\dot{P}&=&-\Gamma_{\!D_s}\,(\delta P+c_{sn}\delta\tilde{n})-\Gamma_{\!S_s}\delta\tilde{T}\,.
\label{eq:2.1Modes}
\end{eqnarray}

\vspace*{0.2in}\noindent Here, from Eq.~\ref{eq:2.5S} with  $\omega_F = {\rm v}_F q$, we have defined 
\begin{equation}
\Omega_{np}^2=\frac{\omega_F^2}{2}\left(\frac{\partial f_p}{\partial P}\right)_\theta\,.
\label{eq:freqsq}
\end{equation}
The decay rates in Eqs.~\ref{eq:2.1Modes} are given in terms of  $\gamma_0=\hbar q^2/m$ by
\begin{equation}
\gamma_\kappa=\gamma_0\,\frac{\tilde{\kappa}_T'}{\tilde{c}_{V_1}}\,;
\hspace{0.1in}\gamma_\eta=\gamma_0\,\frac{4}{3}\tilde{\eta}\,;\hspace{0.1in}\Gamma_{\!D_s}=\gamma_0\,\tilde{D}_s\,;
\hspace{0.1in}\Gamma_{\!S_s}=\gamma_0\,\tilde{S}_s'\,;\hspace{0.1in}\Gamma_{\!P_s}=\gamma_0\,\tilde{P}_s\,C_{\!P_s}\,,
\label{eq:rates1}
\end{equation}
where $\tilde{S}'_s =\tilde{S}_\epsilon-\tilde{D}_s\,\frac{3}{2}\,c_{sn}$ from Eq.~\ref{eq:4.9Ssprime},  $\tilde{P}_s$  given by Eq.~\ref{eq:4.9Ps} and $\tilde{\kappa}'_T$ is given by Eq.~\ref{eq:4.7kappaTprime}.

The coupled equations are solved using the measured initial conditions for $\delta n_\uparrow(q,0)/n_{0\uparrow}$ and $\delta n_\downarrow(q,0)/n_{0\downarrow}$, which determine the nonvanishing initial conditions for $\delta P(q,0)$ from Eq.~\ref{eq:1.13} and for $\delta\tilde{n}(q,0)$ from Eq.~\ref{eq:1.14}. Then,  $c_{sn}\equiv -\delta P(q,0)/\delta\tilde{n}(q,0)$ from Eq.~\ref{eq:1.15}. In thermal and mechanical equilibrium, we have $\delta{T}(q,0)=0$ and $\delta\dot{\tilde{n}}(q,0)=0$.  With these initial conditions, we solve the coupled time dependent equations exactly, using, for example, Mathematica, to obtain the solutions in terms of the fit parameters. Using the general solution, a joint $\chi^2$ fit to $\delta P(q,t)$ and $\delta\tilde{n}(q,t)$ or to $\delta n_{\uparrow}(q,t)/n_{0\uparrow}$ and $\delta n_{\downarrow}(q,t)/n_{0\downarrow}$ determines the fit parameters.  

The fit parameters include the dimensionless transport properties $\tilde{\eta}$, $\tilde{\kappa}_\epsilon$, $\tilde{S}_\epsilon$,  $\tilde{D}_s$ and the isothermal sound frequency $\omega_T=c_T\,q$, which appears in Eq.~\ref{eq:2.1Modes}. $\omega_T$ is adjusted so that the model matches the observed frequency of the first sound mode, which is $\omega_1=c_Sq$ in the long wavelength limit.  The ratio $\omega_T/\omega_F=c_T/{\rm v}_F$ determines the reduced temperature $\theta = T/T_F$, using Eq.~\ref{eq:4.3S} for $c_T$. With $\theta$ determined, the universal scaling functions defined in \S~\ref{sec:thermo} determine all of the additional thermodynamic properties that appear in Eq.~\ref{eq:2.1Modes}. 

\subsubsection{Energy-Spin Seebeck Coefficient}
\label{sec:Seebeck}

A remarkable constraint on $\tilde{S}_\epsilon$ is provided by the third derivative $\delta\dddot{P}(0)$, enabling  measurement of this fundamental energy-spin Seebeck coefficient.
We recall that $\delta\tilde{n}(0)\neq 0$, $\delta\dot{\tilde{n}}(0)=0$, $\delta\tilde{T}(0)=0$, and $\delta P(0)=-c_{sn}\,\delta\tilde{n}(0)$. 
Using these initial conditions in Eqs.~\ref{eq:2.1Modes}, we find that  $\delta\dot{P}(0)=0$ and $\delta\dot{\tilde{T}}(0)=0$, which require $\delta\ddot{P}(0)=0$ and $\delta\ddot{\tilde{T}}(0)=2/3\,\delta\ddot{\tilde{n}}(0)$. Then, 
\begin{equation}
\delta\dddot{P}(0)=-(c_{sn}\,\Gamma_{\!D_s}+2/3\,\Gamma_{\!S_s})\,\delta\ddot{\tilde{n}}(0)\,.
\label{eq:2.7Modes}
\end{equation}
With  $\Gamma_{\!D_s}=\gamma_0\,\tilde{D}_s$ and  $\Gamma_{\!S_s}=\gamma_0\,(\tilde{S}_\epsilon-\tilde{D}_s\,\frac{3}{2}\,c_{sn})$ from Eqs.~\ref{eq:rates1},  we obtain the important result,
\begin{equation}
\delta\dddot{P}(0)=-\frac{2}{3}\,\tilde{S}_\epsilon\,\gamma_0\,\delta\ddot{\tilde{n}}(0)\,.
\label{eq:2.9Modes}
\end{equation} 
Since $\delta\dot{P}(0)=0$ and $\delta\ddot{P}(0)=0$,  $\delta\dddot{P}(0)$ determines the leading time dependence of the $\delta P(q,t)$ data. As $\delta\ddot{\tilde{n}}(0)$ is well-determined from the data for $\delta\tilde{n}(q,t)$, Eq.~\ref{eq:2.9Modes} provides a strong constraint on  $\tilde{S}_\epsilon$ in the fit of the full model to $\delta P(q,t)$ data. As shown in the main text, the energy-spin Seebeck coefficient $\tilde{S}_\epsilon$ obtained by fitting the full model to the data is independent of reduced temperature in the range $T/T_F=0.4-0.5$ with $\tilde{S}_\epsilon\simeq 2.2\,P$. 

Here we show that $\tilde{S}_\epsilon$ can be extracted directly from the data.  
We employ polynomial fits to the $\delta\bar{P}(t)=\delta n_\downarrow/n_{0\downarrow}-\delta n_\uparrow/n_{0\uparrow}$ and $\delta\tilde{n}(t)=\delta n/n_0$ data. For the fit, we use the first $10$ data points, where $0\leq t\leq 1.6$ ms.  As $\delta\dot{\tilde{n}}(0)=0$, we take $\delta\tilde{n}(t)=A_0 + A_2\, t^2/2 + A_3\, t^3/6$. Noting that $\delta\bar{P}(t)=-2/(1-P_0^2)\,\delta P(t)$ from Eq.~\ref{eq:1.14}, we take $\delta P(t)=C_0 + C_3\, t^3/6 + C_4\, t^4/24$ to assure that $\delta\dot{P}(0)=0$ and $\delta\ddot{P}(0)=0$. We find $\tilde{S}_\epsilon$ as a function of $P$ for $T/T_F$ in the range $0.4-0.5$, where we can assume  that $\tilde{S}_\epsilon$ is nearly independent of $T/T_F$, as shown in Fig.~\ref{fig:Seps}A of the main text. 

\begin{figure}
\centering
\includegraphics[width=3.5in]{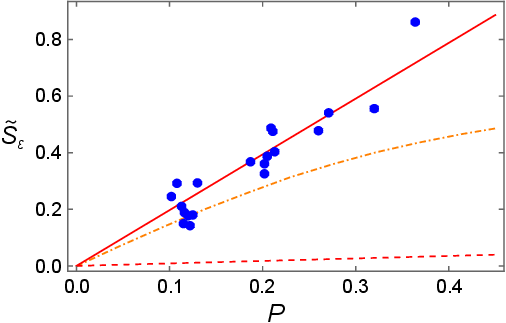}
\caption{\textbf{Energy-spin Seebeck coefficient $S_\epsilon=\tilde{S}_\epsilon\,\frac{n}{T}\,\frac{\hbar}{m}$ versus $P$ from the polynomial method.} Red line:  $\tilde{S}_\epsilon=1.97\,P$, which can be compared to $\tilde{S}_\epsilon=2.20\,P$ obtained from the fits of the full model,  Fig.~\ref{fig:Seps}B of the main text. Red dashed line, kinetic theory limit for $T/T_F=0.45$. Orange dot-dashed curve: Quantum Boltzmann theory prediction for $T/T_F=0.45$, see \S~\ref{sec:Boltzmann}. Error bars are estimated from the $\chi^2$ fit of the polynomials and Eq.~\ref{eq:2.9Modes}.}
\label{fig:SepsPoly}
\end{figure}  

Fig.~\ref{fig:SepsPoly} shows the results from the polynomial fits. The error bars denote the statistical error from the $\chi^2$ fits of the polynomials. We find $\tilde{S}_\epsilon= 1.97\, P$ (red line). The slope is just 10\% smaller than that obtained by fitting the full model. From the two-body (high temperature) kinetic theory limit of \S~\ref{sec:highT}, we find $\tilde{S}_{\epsilon HT}\simeq 5/19\, P\,\tilde{D}_{s0}$ with $\tilde{D}_{s0}\simeq 1.1 (T/T_F)^{3/2}$. For $T/T_F=0.45$,  $\tilde{S}_{\epsilon HT}=0.088\,P$ is shown as the red dashed line. The orange dot-dashed curve shows the quantum Boltzmann theory prediction of \S~\ref{sec:Boltzmann} for $T/T_F=0.45$, where  $\tilde{S}_\epsilon\simeq 1.44\,P$ for $P\leq 0.20$. The predicted slope is consistent in magnitude and sign with the observed slope.
 
We can evaluate the corresponding heat-spin Seebeck coefficient  $\tilde{S}_s$ using Eq.~\ref{eq:tildeSs}, 
\begin{equation}
\tilde{S}_s=\tilde{S}_\epsilon-\tilde{D}_s\,\frac{f_h(P,\theta)}{(\partial f_h/\partial P)_\theta}\simeq\tilde{S}_\epsilon-\tilde{D}_s\,P\,.
\label{eq:SsTilde}
\end{equation}
Here the approximate form follows from $f_h(P=0,\theta)=0$. For $0\leq P\leq 0.4$ and $0.4\leq T/T_F\leq 0.5$, we find from the full equation of state that the coefficient of $\tilde{D}_s\,P$ is unity within 1\%. From the fit to the data of Fig.~\ref{fig:SepsPoly}, we have $\tilde{S}_\epsilon = 1.97\,P$. Using $\tilde{D}_s=2.20$  from the main text,  we find $\tilde{S}_s\simeq -0.23\,P$,  a small negative correlation between the heat and spin currents. A small negative $\tilde{S}_s$ is also obtained in the  kinetic theory limit (see \S~\ref{sec:highT}), where $\tilde{S}_{sHT}=(5/19-39/38)\tilde{D}_{s0}\,P$. For $T/T_F = 0.45$, we find $\tilde{S}_{sHT}\simeq -0.26\,P$. However, as discussed in the main text, the fits of the full model yield $\tilde{S}_\epsilon = 2.2\,P$, which together with $\tilde{D}_s=2.20$ give  $S_s\simeq 0$ for $T/T_F$ in the range $0.4-0.5$, within the measurement uncertainty. 

\subsubsection{First Sound Diffusivity}
\label{sec:D1}

In this section, we show that the first sound diffusivity $D_1$ is independent of $S_\epsilon$ and  $D_s$, and depends only on $\kappa_\epsilon$ and the shear viscosity $\eta$. This important result simplifies the analysis of the data and confirms the consistency of the evolution equations. We begin by looking at the mode structure of the solutions to Eqs.~\ref{eq:2.1Modes}. 

We assume modes of the form $\delta\tilde{n}(q,t)=Ae^{-st}$, $\delta\tilde{T}(q,t)=Be^{-st}$, and $\delta P(q,t)=Ce^{-st}$. Writing the Landau-Placzek parameter as $\epsilon_{LP}=\tilde{c}_{P_1}/\tilde{c}_{V_1}-1=\omega_S^2/\omega_T^2-1$, with $\omega_S=c_S\, q$ the adiabatic sound frequency and $c_S$ the adiabatic sound speed, we find\\ 
\begin{eqnarray}
(s^2+\omega_T^2-s\,\gamma_\eta)\,A+3/2\,(\omega_S^2-\omega_T^2)\,B+\Omega_{np}^2\,C&=&0\nonumber\\
& &\nonumber\\
(2/3\, s + c_{sn}\Gamma_{\!P_s})\,A-(s-\gamma_\kappa)\,B+(s\, C_{T\!P}+\Gamma_{\!P_s})\,C&=&0\nonumber\\
& &\nonumber\\
c_{sn}\Gamma_{\!D_s}\,A+(\Gamma_\epsilon-3/2\, c_{sn}\Gamma_{\!D_s})\,B-(s-\Gamma_{\!D_s})\,C&=&0\,.
\label{eq:2.7Modes}
\end{eqnarray}

\vspace*{0.25in}
The general solution for $\delta P$  takes the form
\begin{equation}
\delta P(q,t)= C_1\,e^{-\Gamma\, t}+C_2\,e^{-c\,t}\,+\,e^{-a\, t}[C_3\cos(bt)+C_4\sin(bt)]
\label{eq:deltaPmodes}
\end{equation}
and similarly for $\delta\tilde{n}(q,t)$ with $C\rightarrow A$ and for $\delta\hat{T}(q,t)$ with $C\rightarrow B$. 

The first sound diffusivity $D_1=\tilde{D}_1\hbar/m$ is determined by the decay rate $a$ of the first sound mode amplitude in Eq.~\ref{eq:deltaPmodes}, where $D_1 q^2=\tilde{D}_1\gamma_0=2a$.  The determinant of the coefficients of $A$, $B$, and $C$  in Eq.~\ref{eq:2.7Modes} is a fourth-order polynomial in $s$ with real coefficients, which vanishes for a nontrivial solution. As Eq.~\ref{eq:deltaPmodes} contains  two decaying heat-spin diffusive modes and a decaying oscillating first sound mode, we assume that the determinant  factors as $(s-\Gamma)(s-c)[(s-a)^2+b^2]$, where $\Gamma$, $a$, $b$, and $c$ are real and positive. Coupled equations for $\Gamma$, $a$, $b$, and $c$ are obtained by comparing the coefficients of $s^n$ in this expression to those in the determinant for $n=3,2,1,0$. The resulting equations are simplified in the long wavelength limit, where the squares of decay rates are small compared to  $\omega_T^2$, $\omega_S^2$, $\Omega_{np}^2\propto\omega_F^2$, and $b^2\simeq\omega_S^2$, yielding 
\begin{equation}
2a=\gamma_\eta+\left(1-\frac{\omega_T^2}{\omega_S^2}\right)\left(\gamma_\kappa +\frac{3}{2}\, c_{sn}\Gamma_{\!P_s}+C_{TP}\Gamma_\epsilon\right)+\frac{2}{3}\,\frac{\Omega_{np}^2}{\omega_S^2}\,\Gamma_\epsilon\,.
\label{eq:9.3S}
\end{equation}

Using Eqs.~\ref{eq:rates1}, it is easy to show that the net contribution of the spin diffusivity vanishes in $\gamma_\kappa +\frac{3}{2}\, c_{sn}\Gamma_{\!P_s}$, yielding
\begin{equation}
2a=\gamma_\eta+\gamma_0\,\left(\frac{1}{\tilde{c}_{V_1}}-\frac{1}{\tilde{c}_{P_1}}\right)\,\tilde{\kappa}_\epsilon+\gamma_0\,\tilde{S}_\epsilon\,Q_\epsilon\,,
\label{eq:9.7S}
\end{equation}
where $\tilde{c}_{V_1}$ ($\tilde{c}_{P_1}$) is the heat capacity per particle at constant volume (pressure) in units of $k_B$ and
\begin{equation}
Q_\epsilon=\frac{4}{15}\frac{1}{f_p}\left[\frac{3}{2}\left(\frac{\partial f_p}{\partial P}\right)_\theta-\theta\,\left(\frac{\partial f_{s_1}}{\partial P}\right)_\theta-f_h\right]
\label{eq:10.3S}
\end{equation}
Here, we have used the universal properties of a unitary Fermi gas, where the pressure $p$ is given by Eq.~\ref{eq:4.1S} and $h$ is given by Eq.~\ref{eq:htheor}. The adiabatic sound speed is given by $mc_S^2=\frac{5}{3}\frac{p}{n}$, which shows that $\frac{\omega_S^2}{\omega_F^2}=\frac{5}{6}\, f_p$. In addition, for the unitary gas, $\left(\frac{1}{\tilde{c}_{V_1}}-\frac{1}{\tilde{c}_{P_1}}\right)=\frac{4}{15}\frac{\theta}{f_p}$~\cite{XinHydroRelax,MZSound}. 
The unitary Fermi gas relation $\frac{3}{2}dp=d{\cal E}$, with $d{\cal E}$ given by Eq.~\ref{eq:1.4Sa} and $n_{\uparrow\downarrow}=\frac{1\pm P}{2}\,n$, then shows that $Q_\epsilon=0$.
Hence, the first sound diffusivity is independent of both the spin diffusivity $D_s$ and the energy-spin Seebeck $S_\epsilon$ (or Peltier $P_\epsilon =TS_\epsilon$) coefficient, which simplifies the analysis of the data.

Finally, we obtain the first sound diffusivity $D_1=\tilde{D}_1\,\hbar/m$, where
\begin{equation}
\tilde{D}_1=\frac{4}{3}\,\tilde{\eta}+\left(\frac{1}{\tilde{c}_{V_1}}-\frac{1}{\tilde{c}_{P_1}}\right)\,\tilde{\kappa}_\epsilon\,.
\label{eq:6.2Modes}
\end{equation}
For a spin-imbalanced unitary Fermi gas, the first sound diffusivity has the usual~\cite{LandauFluids} form in terms of the energy conductivity $\kappa_\epsilon$ at constant $h/T$, substantiating our choice of energy-spin transport coefficients.

We can understand the simple form of Eq.~\ref{eq:6.2Modes}, which is a consequence  of universal thermodynamics for a unitary Fermi gas. Using arguments similar to those of Ref.~\cite{LandauFluids}, the  loss rate of mechanical energy is $\dot{E}=-T_0\,\dot{S}=-2a\,E$, where $\dot{S}$ is the total entropy production rate, given by Eq.~\ref{eq:thermoforces}. For the imbalanced gas, Eq.~\ref{eq:htheor} shows that $h/T=k_B\,f_h(P,\theta)/\theta$ is constant for adiabatic compression, as in the first sound mode. 
In this case, compression preserves $P$, while adiabaticity requires constant entropy per particle $s_1(P,\theta)$, so that $\theta$ is also constant. For constant $h/T$, Eq.~\ref{eq:thermoforces} shows that the total entropy production rate $\dot{S}$ is determined by $-\nabla T\cdot{\mathbf J}_{\epsilon}/T^2=\kappa_\epsilon\,(\nabla T)^2/T^2$, where we omit the volume integrals and $\dot{q}_{fric}$ for brevity. Hence, $\dot{S}$ takes the same form as assumed in the general arguments of Ref.~\cite{LandauFluids} with $\kappa\rightarrow\kappa_\epsilon$, yielding Eq.~\ref{eq:6.2Modes}. This confirms the consistency of the evolution equations, Eq.~\ref{eq:2.1Modes}.

We omit the corresponding results for  $c$ and $\Gamma $ in the long wavelength limit, as they do not provide further insights. However, we note that the decay rates obtained in the long wavelength limit, $2\,a=\tilde{D}_1\,\gamma_0$ for the first sound mode,  $c=\tilde{c}\,\gamma_0$ and $\Gamma=\tilde{\Gamma}\,\gamma_0$ for the heat-spin modes, agree within a few percent with the decay rates obtained from the exact general solution to Eqs.~\ref{eq:2.1Modes} for the same fit parameters. This confirms that the measurements are performed near the long wavelength limit.

\begin{figure}[htb]
\centering
\includegraphics[width=3.5in]{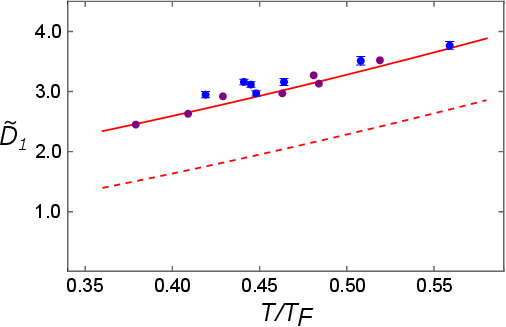}
\caption{\textbf{First sound diffusivity $D_1=\tilde{D}_1\frac{\hbar}{m}$ versus reduced temperature.} $P=0.2$ (blue dots); $P=0$ (purple dots); Red dashed curve, kinetic theory limit Eq.~\ref{eq:D1HTP} for $P=0$. Red-solid curve, Eq.~\ref{eq:D1HTP} for $P=0.2$ with a density shift of $0.82$ added to $\tilde{D}_{1HT0}$. Error bars are estimated from the $\chi^2$ fits of the spin hydrodynamics model using Eq.~\ref{eq:6.2Modes}.}
\label{fig:D1vstheta}
\end{figure} 
  
Measurements of the first sound diffusivity $D_1=\tilde{D}_1\,\hbar/m$ versus reduced temperature $\theta = T/T_F$ are shown in Fig.~\ref{fig:D1vstheta} for fixed polarization in the range $0.187\leq P\leq 0.213$, with an average $\langle P\rangle =0.20$ (blue dots). As a reference, we show data for $P=0$  (purple dots). 
The data can be compared to the high temperature limit, where $1/\tilde{c}_{V_1}-1/\tilde{c}_{P_1}\rightarrow 4/15$ in Eq.~\ref{eq:6.2Modes}. Recalling that $\eta=\tilde{\eta}\hbar n_0$ and $\kappa_\epsilon=\tilde{\kappa}_\epsilon (k_B/m)\hbar n_0$, the kinetic theory limits from \S~\ref{sec:highT} give $\tilde{D}_{1HT0}=7/3\times\,2.77\,\theta^{3/2}$ for $P=0$, which is shown as the red dashed curve. The measured values are comparable to those of previous measurements~\cite{XiangDensityshift,MZSound} and  smaller than the value $\tilde{D}_1\simeq 4.2$  predicted for $T/T_F=0.4$ in Ref.~\cite{EnssTransport}.
  
For $P\neq 0$,  we find $\tilde{D}_1$ from Eq.~\ref{eq:6.2Modes} using the kinetic theory predictions of \S~\ref{sec:highT} and $1/\tilde{c}_{V_1}-1/\tilde{c}_{P_1}\simeq 4/15$,  
\begin{equation}
\tilde{D}_{1HT}(P)= \tilde{D}_{1HT0}\,\frac{1+\left(\frac{8}{35}+\frac{16}{133}\right)\,P^2}{1-P^2}\,,
\label{eq:D1HTP}
\end{equation} 
which gives a small upward shift $\tilde{D}_{1HT}(0.2)\simeq 1.06\,\tilde{D}_{1HT0}$. The solid red curve shows  $\tilde{D}_{1HT}$ versus $T/T_F$ for $P=0.2$ with $\tilde{D}_{1HT0}\rightarrow \tilde{D}_{1HT0}+0.82$, which includes a universal density shift~\cite{XiangDensityshift}. Here the density shift, $0.82$ for $P=0$, was observed in our previous work~\cite{XiangDensityshift}. 

\begin{figure}[htb]
\centering
\includegraphics[width=3.5in]{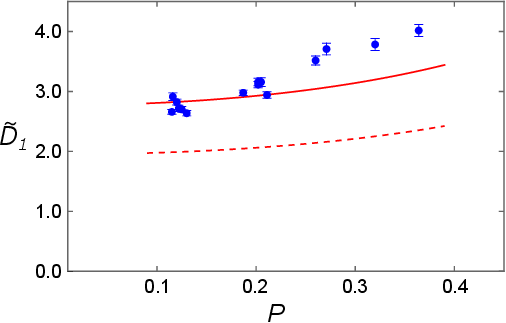}
 \caption{\textbf{First sound diffusivity $D_1=\tilde{D}_1\frac{\hbar}{m}$ versus polarization.} Red dashed curve, kinetic theory limit Eq.~\ref{eq:D1HTP} for $T/T_F=0.45$. Red-solid curve, Eq.~\ref{eq:D1HTP} with a density shift of $0.82$ added to $\tilde{D}_{1HT0}$ for $T/T_F=0.45$. Error bars are estimated from the $\chi^2$ fits of the spin hydrodynamics model using Eq.~\ref{eq:6.2Modes}.}
    \label{fig:D1vsP}
\end{figure}  

Fig.~\ref{fig:D1vsP} shows the first sound diffusivity in units of $\hbar/m$ for $0.4\leq T/T_F\leq0.5$ as a function of polarization $P$. We see that in the normal fluid regime, $\tilde{D}_1$ increases with polarization. The red dashed curve shows the kinetic theory limit, Eq.~\ref{eq:D1HTP} for $\theta =0.45$.  Adding a density shift of $0.82$ to $\tilde{D}_{1HT0}$ in Eq.~\ref{eq:D1HTP}, and assuming the same scaling with $P$, we obtain the solid red curve. For the larger $P$, the data appears to scale more strongly with $P$ than predicted by kinetic theory.


\begin{thebibliography}{45}%
\makeatletter
\providecommand \@ifxundefined [1]{%
 \@ifx{#1\undefined}
}%
\providecommand \@ifnum [1]{%
 \ifnum #1\expandafter \@firstoftwo
 \else \expandafter \@secondoftwo
 \fi
}%
\providecommand \@ifx [1]{%
 \ifx #1\expandafter \@firstoftwo
 \else \expandafter \@secondoftwo
 \fi
}%
\providecommand \natexlab [1]{#1}%
\providecommand \enquote  [1]{``#1''}%
\providecommand \bibnamefont  [1]{#1}%
\providecommand \bibfnamefont [1]{#1}%
\providecommand \citenamefont [1]{#1}%
\providecommand \href@noop [0]{\@secondoftwo}%
\providecommand \href [0]{\begingroup \@sanitize@url \@href}%
\providecommand \@href[1]{\@@startlink{#1}\@@href}%
\providecommand \@@href[1]{\endgroup#1\@@endlink}%
\providecommand \@sanitize@url [0]{\catcode `\\12\catcode `\$12\catcode
  `\&12\catcode `\#12\catcode `\^12\catcode `\_12\catcode `\%12\relax}%
\providecommand \@@startlink[1]{}%
\providecommand \@@endlink[0]{}%
\providecommand \url  [0]{\begingroup\@sanitize@url \@url }%
\providecommand \@url [1]{\endgroup\@href {#1}{\urlprefix }}%
\providecommand \urlprefix  [0]{URL }%
\providecommand \Eprint [0]{\href }%
\providecommand \doibase [0]{https://doi.org/}%
\providecommand \selectlanguage [0]{\@gobble}%
\providecommand \bibinfo  [0]{\@secondoftwo}%
\providecommand \bibfield  [0]{\@secondoftwo}%
\providecommand \translation [1]{[#1]}%
\providecommand \BibitemOpen [0]{}%
\providecommand \bibitemStop [0]{}%
\providecommand \bibitemNoStop [0]{.\EOS\space}%
\providecommand \EOS [0]{\spacefactor3000\relax}%
\providecommand \BibitemShut  [1]{\csname bibitem#1\endcsname}%
\let\auto@bib@innerbib\@empty
\bibitem [{\citenamefont {{O'Hara}}\ \emph {et~al.}(2002)\citenamefont
  {{O'Hara}}, \citenamefont {{Hemmer}}, \citenamefont {{Gehm}}, \citenamefont
  {{Granade}},\ and\ \citenamefont {{Thomas}}}]{OHaraScience}%
  \BibitemOpen
  \bibfield  {author} {\bibinfo {author} {\bibfnamefont {K.~M.}\ \bibnamefont
  {{O'Hara}}}, \bibinfo {author} {\bibfnamefont {S.~L.}\ \bibnamefont
  {{Hemmer}}}, \bibinfo {author} {\bibfnamefont {M.~E.}\ \bibnamefont
  {{Gehm}}}, \bibinfo {author} {\bibfnamefont {S.~R.}\ \bibnamefont
  {{Granade}}},\ and\ \bibinfo {author} {\bibfnamefont {J.~E.}\ \bibnamefont
  {{Thomas}}},\ }\bibfield  {title} {\bibinfo {title} {Observation of a
  strongly interacting degenerate {Fermi} gas of atoms},\ }\href@noop {}
  {\bibfield  {journal} {\bibinfo  {journal} {Science}\ }\textbf {\bibinfo
  {volume} {298}},\ \bibinfo {pages} {2179} (\bibinfo {year}
  {2002})}\BibitemShut {NoStop}%
\bibitem [{\citenamefont {{Lucas}}\ and\ \citenamefont
  {{Fong}}(2018)}]{GrapheneHydro2018}%
  \BibitemOpen
  \bibfield  {author} {\bibinfo {author} {\bibfnamefont {A.}~\bibnamefont
  {{Lucas}}}\ and\ \bibinfo {author} {\bibfnamefont {K.~C.}\ \bibnamefont
  {{Fong}}},\ }\bibfield  {title} {\bibinfo {title} {Hydrodynamics of electrons
  in graphene},\ }\href@noop {} {\bibfield  {journal} {\bibinfo  {journal} {J.
  Phys.: Condens. Matter}\ }\textbf {\bibinfo {volume} {30}},\ \bibinfo {pages}
  {053001} (\bibinfo {year} {2018})}\BibitemShut {NoStop}%
\bibitem [{\citenamefont {{Adams}}\ \emph {et~al.}(2012)\citenamefont
  {{Adams}}, \citenamefont {{Carr}}, \citenamefont {{Sch\"{a}fer}},
  \citenamefont {{Steinberg}},\ and\ \citenamefont {{Thomas}}}]{NJPReview}%
  \BibitemOpen
  \bibfield  {author} {\bibinfo {author} {\bibfnamefont {A.}~\bibnamefont
  {{Adams}}}, \bibinfo {author} {\bibfnamefont {L.~D.}\ \bibnamefont {{Carr}}},
  \bibinfo {author} {\bibfnamefont {T.}~\bibnamefont {{Sch\"{a}fer}}}, \bibinfo
  {author} {\bibfnamefont {P.}~\bibnamefont {{Steinberg}}},\ and\ \bibinfo
  {author} {\bibfnamefont {J.~E.}\ \bibnamefont {{Thomas}}},\ }\bibfield
  {title} {\bibinfo {title} {Strongly correlated quantum fluids: ultracold
  quantum gases, quantum chromodynamic plasmas and holographic duality},\
  }\href@noop {} {\bibfield  {journal} {\bibinfo  {journal} {New J. Phys.}\
  }\textbf {\bibinfo {volume} {14}},\ \bibinfo {pages} {115009} (\bibinfo
  {year} {2012})}\BibitemShut {NoStop}%
\bibitem [{\citenamefont {{Bloch}}\ \emph {et~al.}(2012)\citenamefont
  {{Bloch}}, \citenamefont {{Dalibard}},\ and\ \citenamefont
  {{Nascimb\`ene}}}]{BlochReview}%
  \BibitemOpen
  \bibfield  {author} {\bibinfo {author} {\bibfnamefont {I.}~\bibnamefont
  {{Bloch}}}, \bibinfo {author} {\bibfnamefont {J.}~\bibnamefont
  {{Dalibard}}},\ and\ \bibinfo {author} {\bibfnamefont {S.}~\bibnamefont
  {{Nascimb\`ene}}},\ }\bibfield  {title} {\bibinfo {title} {Quantum
  simulations with ultracold quantum gases},\ }\href@noop {} {\bibfield
  {journal} {\bibinfo  {journal} {Nat. Phys.}\ }\textbf {\bibinfo {volume}
  {8}},\ \bibinfo {pages} {267} (\bibinfo {year} {2012})}\BibitemShut {NoStop}%
\bibitem [{\citenamefont {{Strinati}}\ \emph {et~al.}(2018)\citenamefont
  {{Strinati}}, \citenamefont {{Pieri}}, \citenamefont {{R\"{o}pke}},
  \citenamefont {{Schuck}},\ and\ \citenamefont {{Urban}}}]{UrbanReview}%
  \BibitemOpen
  \bibfield  {author} {\bibinfo {author} {\bibfnamefont {G.~C.}\ \bibnamefont
  {{Strinati}}}, \bibinfo {author} {\bibfnamefont {P.}~\bibnamefont {{Pieri}}},
  \bibinfo {author} {\bibfnamefont {G.}~\bibnamefont {{R\"{o}pke}}}, \bibinfo
  {author} {\bibfnamefont {P.}~\bibnamefont {{Schuck}}},\ and\ \bibinfo
  {author} {\bibfnamefont {M.}~\bibnamefont {{Urban}}},\ }\bibfield  {title}
  {\bibinfo {title} {The {BCS-BEC} crossover: From ultra-cold {Fermi} gases to
  nuclear systems},\ }\href@noop {} {\bibfield  {journal} {\bibinfo  {journal}
  {Phys. Rep.}\ }\textbf {\bibinfo {volume} {738}},\ \bibinfo {pages} {1}
  (\bibinfo {year} {2018})}\BibitemShut {NoStop}%
\bibitem [{\citenamefont {{The Star Collaboration}}(2017)}]{StarCollab}%
  \BibitemOpen
  \bibfield  {author} {\bibinfo {author} {\bibnamefont {{The Star
  Collaboration}}},\ }\bibfield  {title} {\bibinfo {title} {Global $\lambda$
  hyperon polarization in nuclear collisions},\ }\href@noop {} {\bibfield
  {journal} {\bibinfo  {journal} {Nature}\ }\textbf {\bibinfo {volume} {548}},\
  \bibinfo {pages} {62} (\bibinfo {year} {2017})}\BibitemShut {NoStop}%
\bibitem [{\citenamefont {{Becattini}}\ \emph {et~al.}(2024)\citenamefont
  {{Becattini}}, \citenamefont {{Buzzegoli}}, \citenamefont {{Niida}},
  \citenamefont {{Pu}}, \citenamefont {{Tang}},\ and\ \citenamefont
  {{Wang}}}]{SpinHydroRelativistic}%
  \BibitemOpen
  \bibfield  {author} {\bibinfo {author} {\bibfnamefont {F.}~\bibnamefont
  {{Becattini}}}, \bibinfo {author} {\bibfnamefont {M.}~\bibnamefont
  {{Buzzegoli}}}, \bibinfo {author} {\bibfnamefont {T.}~\bibnamefont
  {{Niida}}}, \bibinfo {author} {\bibfnamefont {S.}~\bibnamefont {{Pu}}},
  \bibinfo {author} {\bibfnamefont {A.-H.}\ \bibnamefont {{Tang}}},\ and\
  \bibinfo {author} {\bibfnamefont {Q.}~\bibnamefont {{Wang}}},\ }\bibfield
  {title} {\bibinfo {title} {Spin polarization in relativistic heavy-ion
  collisions},\ }\href@noop {} {\bibfield  {journal} {\bibinfo  {journal} {Int.
  J. Mod. Phys. E}\ }\textbf {\bibinfo {volume} {33}},\ \bibinfo {pages}
  {2430006} (\bibinfo {year} {2024})}\BibitemShut {NoStop}%
\bibitem [{\citenamefont {Sapna}\ \emph {et~al.}(2025)\citenamefont {Sapna},
  \citenamefont {Singh},\ and\ \citenamefont {Wagner}}]{DissHydroSpinPol}%
  \BibitemOpen
  \bibfield  {author} {\bibinfo {author} {\bibnamefont {Sapna}}, \bibinfo
  {author} {\bibfnamefont {S.~K.}\ \bibnamefont {Singh}},\ and\ \bibinfo
  {author} {\bibfnamefont {D.}~\bibnamefont {Wagner}},\ }\bibfield  {title}
  {\bibinfo {title} {Spin polarization of $\mathrm{\ensuremath{\Lambda}}$
  hyperons from dissipative spin hydrodynamics},\ }\href@noop {} {\bibfield
  {journal} {\bibinfo  {journal} {Phys. Rev. C}\ }\textbf {\bibinfo {volume}
  {112}},\ \bibinfo {pages} {054902} (\bibinfo {year} {2025})}\BibitemShut
  {NoStop}%
\bibitem [{\citenamefont {{Bauer}}\ \emph {et~al.}(2012)\citenamefont
  {{Bauer}}, \citenamefont {{Saitoh}},\ and\ \citenamefont {{Van
  Wees}}}]{SpinCal2012}%
  \BibitemOpen
  \bibfield  {author} {\bibinfo {author} {\bibfnamefont {G.}~\bibnamefont
  {{Bauer}}}, \bibinfo {author} {\bibfnamefont {E.}~\bibnamefont {{Saitoh}}},\
  and\ \bibinfo {author} {\bibfnamefont {B.}~\bibnamefont {{Van Wees}}},\
  }\bibfield  {title} {\bibinfo {title} {Spin caloritronics},\ }\href@noop {}
  {\bibfield  {journal} {\bibinfo  {journal} {Nat. Mater.}\ }\textbf {\bibinfo
  {volume} {11}},\ \bibinfo {pages} {391} (\bibinfo {year} {2012})}\BibitemShut
  {NoStop}%
\bibitem [{\citenamefont {{Adachi}}\ \emph {et~al.}(2013)\citenamefont
  {{Adachi}}, \citenamefont {{Uchida}}, \citenamefont {{Saitoh}},\ and\
  \citenamefont {{Maekawa}}}]{AdachiSpinSeebeck}%
  \BibitemOpen
  \bibfield  {author} {\bibinfo {author} {\bibfnamefont {H.}~\bibnamefont
  {{Adachi}}}, \bibinfo {author} {\bibfnamefont {K.}~\bibnamefont {{Uchida}}},
  \bibinfo {author} {\bibfnamefont {E.}~\bibnamefont {{Saitoh}}},\ and\
  \bibinfo {author} {\bibfnamefont {S.}~\bibnamefont {{Maekawa}}},\ }\bibfield
  {title} {\bibinfo {title} {Theory of the spin {Seebeck} effect},\ }\href@noop
  {} {\bibfield  {journal} {\bibinfo  {journal} {Rep. Prog. Phys.}\ }\textbf
  {\bibinfo {volume} {76}},\ \bibinfo {pages} {036501} (\bibinfo {year}
  {2013})}\BibitemShut {NoStop}%
\bibitem [{\citenamefont {{Uchida}}\ \emph {et~al.}(2008)\citenamefont
  {{Uchida}}, \citenamefont {{Takahashi}}, \citenamefont {{Harii}},
  \citenamefont {{Ieda}}, \citenamefont {{Koshibae}}, \citenamefont {{Ando}},
  \citenamefont {{Maekawa}},\ and\ \citenamefont
  {{Saitoh}}}]{UchidaObsSeebeck}%
  \BibitemOpen
  \bibfield  {author} {\bibinfo {author} {\bibfnamefont {K.}~\bibnamefont
  {{Uchida}}}, \bibinfo {author} {\bibfnamefont {S.}~\bibnamefont
  {{Takahashi}}}, \bibinfo {author} {\bibfnamefont {K.}~\bibnamefont
  {{Harii}}}, \bibinfo {author} {\bibfnamefont {J.}~\bibnamefont {{Ieda}}},
  \bibinfo {author} {\bibfnamefont {W.}~\bibnamefont {{Koshibae}}}, \bibinfo
  {author} {\bibfnamefont {K.}~\bibnamefont {{Ando}}}, \bibinfo {author}
  {\bibfnamefont {S.}~\bibnamefont {{Maekawa}}},\ and\ \bibinfo {author}
  {\bibfnamefont {E.}~\bibnamefont {{Saitoh}}},\ }\bibfield  {title} {\bibinfo
  {title} {Observation of the spin seebeck effect},\ }\href@noop {} {\bibfield
  {journal} {\bibinfo  {journal} {Nature}\ }\textbf {\bibinfo {volume} {455}},\
  \bibinfo {pages} {778} (\bibinfo {year} {2008})}\BibitemShut {NoStop}%
\bibitem [{\citenamefont {{Yang}}\ \emph {et~al.}(2023)\citenamefont {{Yang}},
  \citenamefont {{Sang}}, \citenamefont {{Zhang}}, \citenamefont {{Hamilton}},
  \citenamefont {{Fuhrer}},\ and\ \citenamefont {{Wang}}}]{SpinThermoElec}%
  \BibitemOpen
  \bibfield  {author} {\bibinfo {author} {\bibfnamefont {G.}~\bibnamefont
  {{Yang}}}, \bibinfo {author} {\bibfnamefont {L.}~\bibnamefont {{Sang}}},
  \bibinfo {author} {\bibfnamefont {N.}~\bibnamefont {{Zhang}}, \bibfnamefont
  {C.and~{Ye}}}, \bibinfo {author} {\bibfnamefont {A.}~\bibnamefont
  {{Hamilton}}}, \bibinfo {author} {\bibfnamefont {M.~S.}\ \bibnamefont
  {{Fuhrer}}},\ and\ \bibinfo {author} {\bibfnamefont {X.}~\bibnamefont
  {{Wang}}},\ }\bibfield  {title} {\bibinfo {title} {The role of spin in
  thermoelectricity},\ }\href@noop {} {\bibfield  {journal} {\bibinfo
  {journal} {Nat. Rev. Phys.}\ }\textbf {\bibinfo {volume} {5}},\ \bibinfo
  {pages} {466} (\bibinfo {year} {2023})}\BibitemShut {NoStop}%
\bibitem [{\citenamefont {{Kim}}\ and\ \citenamefont
  {{Huse}}(2012)}]{KimHuseSpinDiff}%
  \BibitemOpen
  \bibfield  {author} {\bibinfo {author} {\bibfnamefont {H.}~\bibnamefont
  {{Kim}}}\ and\ \bibinfo {author} {\bibfnamefont {D.~A.}\ \bibnamefont
  {{Huse}}},\ }\bibfield  {title} {\bibinfo {title} {Heat and spin transport in
  a cold atomic {Fermi} gas},\ }\href@noop {} {\bibfield  {journal} {\bibinfo
  {journal} {Phys. Rev. A}\ }\textbf {\bibinfo {volume} {86}},\ \bibinfo
  {pages} {053607} (\bibinfo {year} {2012})}\BibitemShut {NoStop}%
\bibitem [{\citenamefont {{Wong}}\ \emph {et~al.}(2012)\citenamefont {{Wong}},
  \citenamefont {{Stoof}},\ and\ \citenamefont
  {{Duine}}}]{DuineStoofSpinSeebeckFG}%
  \BibitemOpen
  \bibfield  {author} {\bibinfo {author} {\bibfnamefont {C.~H.}\ \bibnamefont
  {{Wong}}}, \bibinfo {author} {\bibfnamefont {H.~T.~C.}\ \bibnamefont
  {{Stoof}}},\ and\ \bibinfo {author} {\bibfnamefont {R.~A.}\ \bibnamefont
  {{Duine}}},\ }\bibfield  {title} {\bibinfo {title} {Spin-seebeck effect in a
  strongly interacting {Fermi} gas},\ }\href@noop {} {\bibfield  {journal}
  {\bibinfo  {journal} {Phys. Rev. A}\ }\textbf {\bibinfo {volume} {85}},\
  \bibinfo {pages} {063613} (\bibinfo {year} {2012})}\BibitemShut {NoStop}%
\bibitem [{\citenamefont {{Sommer}}\ \emph {et~al.}(2011)\citenamefont
  {{Sommer}}, \citenamefont {{Ku}}, \citenamefont {{Roati}},\ and\
  \citenamefont {{Zwierlein}}}]{SommerSpinDiff}%
  \BibitemOpen
  \bibfield  {author} {\bibinfo {author} {\bibfnamefont {A.}~\bibnamefont
  {{Sommer}}}, \bibinfo {author} {\bibfnamefont {M.}~\bibnamefont {{Ku}}},
  \bibinfo {author} {\bibfnamefont {G.}~\bibnamefont {{Roati}}},\ and\ \bibinfo
  {author} {\bibfnamefont {M.~W.}\ \bibnamefont {{Zwierlein}}},\ }\bibfield
  {title} {\bibinfo {title} {Universal spin transport in a strongly interacting
  {Fermi} gas},\ }\href@noop {} {\bibfield  {journal} {\bibinfo  {journal}
  {Nature}\ }\textbf {\bibinfo {volume} {472}},\ \bibinfo {pages} {201}
  (\bibinfo {year} {2011})}\BibitemShut {NoStop}%
\bibitem [{\citenamefont {{Ho}}(2004)}]{HoUniversalThermo}%
  \BibitemOpen
  \bibfield  {author} {\bibinfo {author} {\bibfnamefont {T.~L.}\ \bibnamefont
  {{Ho}}},\ }\bibfield  {title} {\bibinfo {title} {Universal thermodynamics of
  degenerate quantum gases in the unitarity limit},\ }\href@noop {} {\bibfield
  {journal} {\bibinfo  {journal} {Phys. Rev. Lett.}\ }\textbf {\bibinfo
  {volume} {92}},\ \bibinfo {pages} {090402} (\bibinfo {year}
  {2004})}\BibitemShut {NoStop}%
\bibitem [{\citenamefont {Navon}\ \emph {et~al.}(2021)\citenamefont {Navon},
  \citenamefont {Smith},\ and\ \citenamefont {Hadzibabic}}]{NavonBox}%
  \BibitemOpen
  \bibfield  {author} {\bibinfo {author} {\bibfnamefont {N.}~\bibnamefont
  {Navon}}, \bibinfo {author} {\bibfnamefont {R.}~\bibnamefont {Smith}},\ and\
  \bibinfo {author} {\bibfnamefont {Z.}~\bibnamefont {Hadzibabic}},\ }\bibfield
   {title} {\bibinfo {title} {Quantum gases in optical boxes},\ }\href@noop {}
  {\bibfield  {journal} {\bibinfo  {journal} {Nat. Phys.}\ }\textbf {\bibinfo
  {volume} {17}},\ \bibinfo {pages} {1334–1341} (\bibinfo {year}
  {2021})}\BibitemShut {NoStop}%
\bibitem [{\citenamefont {{Bartenstein}}\ \emph {et~al.}(2005)\citenamefont
  {{Bartenstein}}, \citenamefont {{Altmeyer}}, \citenamefont {{Riedl}},
  \citenamefont {{Geursen}}, \citenamefont {{Jochim}}, \citenamefont {{Chin}},
  \citenamefont {{Denschlag}}, \citenamefont {R.}, \citenamefont {A.},
  \citenamefont {E.}, \citenamefont {J.},\ and\ \citenamefont
  {{Julienne}}}]{BartensteinFeshbach}%
  \BibitemOpen
  \bibfield  {author} {\bibinfo {author} {\bibfnamefont {M.}~\bibnamefont
  {{Bartenstein}}}, \bibinfo {author} {\bibfnamefont {A.}~\bibnamefont
  {{Altmeyer}}}, \bibinfo {author} {\bibfnamefont {S.}~\bibnamefont {{Riedl}}},
  \bibinfo {author} {\bibfnamefont {R.}~\bibnamefont {{Geursen}}}, \bibinfo
  {author} {\bibfnamefont {S.}~\bibnamefont {{Jochim}}}, \bibinfo {author}
  {\bibfnamefont {C.}~\bibnamefont {{Chin}}}, \bibinfo {author} {\bibfnamefont
  {J.~H.}\ \bibnamefont {{Denschlag}}}, \bibinfo {author} {\bibfnamefont
  {G.}~\bibnamefont {R.}}, \bibinfo {author} {\bibfnamefont {S.}~\bibnamefont
  {A.}}, \bibinfo {author} {\bibfnamefont {T.}~\bibnamefont {E.}}, \bibinfo
  {author} {\bibfnamefont {W.~C.}\ \bibnamefont {J.}},\ and\ \bibinfo {author}
  {\bibfnamefont {P.~S.}\ \bibnamefont {{Julienne}}},\ }\bibfield  {title}
  {\bibinfo {title} {Precise determination of $^6$\mbox{Li} cold collision
  parameters by radio-frequency spectroscopy on weakly bound molecules},\
  }\href@noop {} {\bibfield  {journal} {\bibinfo  {journal} {Phys. Rev. Lett.}\
  }\textbf {\bibinfo {volume} {94}},\ \bibinfo {pages} {103201} (\bibinfo
  {year} {2005})}\BibitemShut {NoStop}%
\bibitem [{\citenamefont {{Z\"urn}}\ \emph {et~al.}(2013)\citenamefont
  {{Z\"urn}}, \citenamefont {{Lompe}}, \citenamefont {{Wenz}}, \citenamefont
  {{Jochim}}, \citenamefont {{Julienne}},\ and\ \citenamefont
  {{Hutson}}}]{JochimPreciseFeshbach}%
  \BibitemOpen
  \bibfield  {author} {\bibinfo {author} {\bibfnamefont {G.}~\bibnamefont
  {{Z\"urn}}}, \bibinfo {author} {\bibfnamefont {T.}~\bibnamefont {{Lompe}}},
  \bibinfo {author} {\bibfnamefont {A.~N.}\ \bibnamefont {{Wenz}}}, \bibinfo
  {author} {\bibfnamefont {S.}~\bibnamefont {{Jochim}}}, \bibinfo {author}
  {\bibfnamefont {P.~S.}\ \bibnamefont {{Julienne}}},\ and\ \bibinfo {author}
  {\bibfnamefont {J.~M.}\ \bibnamefont {{Hutson}}},\ }\bibfield  {title}
  {\bibinfo {title} {Precise characterization of $^{6}\mathrm{Li}$ {Feshbach}
  resonances using trap-sideband-resolved rf spectroscopy of weakly bound
  molecules},\ }\href@noop {} {\bibfield  {journal} {\bibinfo  {journal} {Phys.
  Rev. Lett.}\ }\textbf {\bibinfo {volume} {110}},\ \bibinfo {pages} {135301}
  (\bibinfo {year} {2013})}\BibitemShut {NoStop}%
\bibitem [{\citenamefont {{Baird}}\ \emph {et~al.}(2019)\citenamefont
  {{Baird}}, \citenamefont {{Wang}}, \citenamefont {{Roof}},\ and\
  \citenamefont {{Thomas}}}]{LorinLinearHydro}%
  \BibitemOpen
  \bibfield  {author} {\bibinfo {author} {\bibfnamefont {L.}~\bibnamefont
  {{Baird}}}, \bibinfo {author} {\bibfnamefont {X.}~\bibnamefont {{Wang}}},
  \bibinfo {author} {\bibfnamefont {S.}~\bibnamefont {{Roof}}},\ and\ \bibinfo
  {author} {\bibfnamefont {J.~E.}\ \bibnamefont {{Thomas}}},\ }\bibfield
  {title} {\bibinfo {title} {Measuring the hydrodynamic linear response of a
  unitary {Fermi} gas},\ }\href@noop {} {\bibfield  {journal} {\bibinfo
  {journal} {Phys. Rev. Lett.}\ }\textbf {\bibinfo {volume} {123}},\ \bibinfo
  {pages} {160402} (\bibinfo {year} {2019})}\BibitemShut {NoStop}%
\bibitem [{\citenamefont {{Wang}}\ \emph {et~al.}(2022)\citenamefont {{Wang}},
  \citenamefont {{Li}}, \citenamefont {{Arakelyan}},\ and\ \citenamefont
  {{Thomas}}}]{XinHydroRelax}%
  \BibitemOpen
  \bibfield  {author} {\bibinfo {author} {\bibfnamefont {X.}~\bibnamefont
  {{Wang}}}, \bibinfo {author} {\bibfnamefont {X.}~\bibnamefont {{Li}}},
  \bibinfo {author} {\bibfnamefont {I.}~\bibnamefont {{Arakelyan}}},\ and\
  \bibinfo {author} {\bibfnamefont {J.~E.}\ \bibnamefont {{Thomas}}},\
  }\bibfield  {title} {\bibinfo {title} {Hydrodynamic relaxation in a strongly
  interacting {Fermi} gas},\ }\href@noop {} {\bibfield  {journal} {\bibinfo
  {journal} {Phys. Rev. Lett.}\ }\textbf {\bibinfo {volume} {128}},\ \bibinfo
  {pages} {090402} (\bibinfo {year} {2022})}\BibitemShut {NoStop}%
\bibitem [{Sup()}]{SupportOnline}%
  \BibitemOpen
  \href@noop {} {}\bibinfo {note} {See the Supplemental Text for discussions of
  the linearized hydrodynamic equations, the definitions of the heat, energy,
  and spin currents, and the choice of transport coefficients.}\BibitemShut
  {Stop}%
\bibitem [{\citenamefont {Landau}\ and\ \citenamefont
  {Lifshitz}(1959)}]{LandauFluids}%
  \BibitemOpen
  \bibfield  {author} {\bibinfo {author} {\bibfnamefont {L.~D.}\ \bibnamefont
  {Landau}}\ and\ \bibinfo {author} {\bibfnamefont {E.~M.}\ \bibnamefont
  {Lifshitz}},\ }\href@noop {} {\emph {\bibinfo {title} {Fluid Dynamics, Course
  of Theoretical Physics Vol. VI}}}\ (\bibinfo  {publisher} {Pergamon Press,
  Oxford},\ \bibinfo {year} {1959})\BibitemShut {NoStop}%
\bibitem [{Bul()}]{Bulk}%
  \BibitemOpen
  \href@noop {} {}\bibinfo {note} {For a unitary Fermi gas, the bulk viscosity
  $\xi_B$ has been measured~\cite{ElliottScaleInv} and found to be negligible
  compared to the shear viscosity, consistent with predictions that $\xi_B=0$
  for scale invariant
  systems~\cite{SonBulkViscosity,StringariBulk}.}\BibitemShut {Stop}%
\bibitem [{\citenamefont {{Elliott}}\ \emph {et~al.}(2014)\citenamefont
  {{Elliott}}, \citenamefont {{Joseph}},\ and\ \citenamefont
  {{Thomas}}}]{ElliottScaleInv}%
  \BibitemOpen
  \bibfield  {author} {\bibinfo {author} {\bibfnamefont {E.}~\bibnamefont
  {{Elliott}}}, \bibinfo {author} {\bibfnamefont {J.~A.}\ \bibnamefont
  {{Joseph}}},\ and\ \bibinfo {author} {\bibfnamefont {J.~E.}\ \bibnamefont
  {{Thomas}}},\ }\bibfield  {title} {\bibinfo {title} {Observation of conformal
  symmetry breaking and scale invariance in expanding {Fermi} gases},\
  }\href@noop {} {\bibfield  {journal} {\bibinfo  {journal} {Phys. Rev. Lett.}\
  }\textbf {\bibinfo {volume} {112}},\ \bibinfo {pages} {040405} (\bibinfo
  {year} {2014})}\BibitemShut {NoStop}%
\bibitem [{\citenamefont {{Son}}(2007)}]{SonBulkViscosity}%
  \BibitemOpen
  \bibfield  {author} {\bibinfo {author} {\bibfnamefont {D.~T.}\ \bibnamefont
  {{Son}}},\ }\bibfield  {title} {\bibinfo {title} {Vanishing bulk viscosities
  and conformal invariance of the unitary {Fermi} gas},\ }\href@noop {}
  {\bibfield  {journal} {\bibinfo  {journal} {Phys. Rev. Lett.}\ }\textbf
  {\bibinfo {volume} {98}},\ \bibinfo {pages} {020604} (\bibinfo {year}
  {2007})}\BibitemShut {NoStop}%
\bibitem [{\citenamefont {{Hou}}\ \emph {et~al.}(2013)\citenamefont {{Hou}},
  \citenamefont {{Pitaevskii}},\ and\ \citenamefont
  {{Stringari}}}]{StringariBulk}%
  \BibitemOpen
  \bibfield  {author} {\bibinfo {author} {\bibfnamefont {Y.~H.}\ \bibnamefont
  {{Hou}}}, \bibinfo {author} {\bibfnamefont {L.~P.}\ \bibnamefont
  {{Pitaevskii}}},\ and\ \bibinfo {author} {\bibfnamefont {S.}~\bibnamefont
  {{Stringari}}},\ }\bibfield  {title} {\bibinfo {title} {Scaling solutions of
  the two-fluid hydrodynamic equations in a harmonically trapped gas at
  unitarity},\ }\href@noop {} {\bibfield  {journal} {\bibinfo  {journal} {Phys.
  Rev. A}\ }\textbf {\bibinfo {volume} {87}},\ \bibinfo {pages} {033620}
  (\bibinfo {year} {2013})}\BibitemShut {NoStop}%
\bibitem [{\citenamefont {Frank}\ \emph {et~al.}(2018)\citenamefont {Frank},
  \citenamefont {Lang},\ and\ \citenamefont
  {Zwerger}}]{FrankLangZwergPhaseDiag}%
  \BibitemOpen
  \bibfield  {author} {\bibinfo {author} {\bibfnamefont {B.}~\bibnamefont
  {Frank}}, \bibinfo {author} {\bibfnamefont {J.}~\bibnamefont {Lang}},\ and\
  \bibinfo {author} {\bibfnamefont {W.}~\bibnamefont {Zwerger}},\ }\bibfield
  {title} {\bibinfo {title} {Universal phase diagram and scaling functions of
  imbalanced {Fermi} gases},\ }\href@noop {} {\bibfield  {journal} {\bibinfo
  {journal} {J. Exp. Theor. Phys.}\ }\textbf {\bibinfo {volume} {127}},\
  \bibinfo {pages} {812–825} (\bibinfo {year} {2018})}\BibitemShut {NoStop}%
\bibitem [{\citenamefont {{Ku}}\ \emph {et~al.}(2012)\citenamefont {{Ku}},
  \citenamefont {{Sommer}}, \citenamefont {{Cheuk}},\ and\ \citenamefont
  {{Zwierlein}}}]{KuThermo}%
  \BibitemOpen
  \bibfield  {author} {\bibinfo {author} {\bibfnamefont {M.~J.}\ \bibnamefont
  {{Ku}}}, \bibinfo {author} {\bibfnamefont {A.~T.}\ \bibnamefont {{Sommer}}},
  \bibinfo {author} {\bibfnamefont {L.~W.}\ \bibnamefont {{Cheuk}}},\ and\
  \bibinfo {author} {\bibfnamefont {M.~W.}\ \bibnamefont {{Zwierlein}}},\
  }\bibfield  {title} {\bibinfo {title} {Revealing the superfluid lambda
  transition in the universal thermodynamics of a unitary {Fermi} gas},\
  }\href@noop {} {\bibfield  {journal} {\bibinfo  {journal} {Science}\ }\textbf
  {\bibinfo {volume} {335}},\ \bibinfo {pages} {563} (\bibinfo {year}
  {2012})}\BibitemShut {NoStop}%
\bibitem [{\citenamefont {{Bruun}}(2011)}]{BruunSpinDiff}%
  \BibitemOpen
  \bibfield  {author} {\bibinfo {author} {\bibfnamefont {G.~M.}\ \bibnamefont
  {{Bruun}}},\ }\bibfield  {title} {\bibinfo {title} {Spin diffusion in {Fermi}
  gases},\ }\href@noop {} {\bibfield  {journal} {\bibinfo  {journal} {New J.
  Phys.}\ }\textbf {\bibinfo {volume} {13}},\ \bibinfo {pages} {035005}
  (\bibinfo {year} {2011})}\BibitemShut {NoStop}%
\bibitem [{\citenamefont {Enss}(2012)}]{Enss2012}%
  \BibitemOpen
  \bibfield  {author} {\bibinfo {author} {\bibfnamefont {T.}~\bibnamefont
  {Enss}},\ }\bibfield  {title} {\bibinfo {title} {Quantum critical transport
  in the unitary {Fermi} gas},\ }\href@noop {} {\bibfield  {journal} {\bibinfo
  {journal} {Phys. Rev. A}\ }\textbf {\bibinfo {volume} {86}},\ \bibinfo
  {pages} {013616} (\bibinfo {year} {2012})}\BibitemShut {NoStop}%
\bibitem [{\citenamefont {{Enss}}\ and\ \citenamefont
  {{Haussmann}}(2012)}]{EnssSpinDiff}%
  \BibitemOpen
  \bibfield  {author} {\bibinfo {author} {\bibfnamefont {T.}~\bibnamefont
  {{Enss}}}\ and\ \bibinfo {author} {\bibfnamefont {R.}~\bibnamefont
  {{Haussmann}}},\ }\bibfield  {title} {\bibinfo {title} {Quantum mechanical
  limitations to spin diffusion in the unitary {Fermi} gas},\ }\href@noop {}
  {\bibfield  {journal} {\bibinfo  {journal} {Phys. Rev. Lett.}\ }\textbf
  {\bibinfo {volume} {109}},\ \bibinfo {pages} {195303} (\bibinfo {year}
  {2012})}\BibitemShut {NoStop}%
\bibitem [{\citenamefont {Sch\"{a}fer}(2016)}]{SchaferSpinDiff}%
  \BibitemOpen
  \bibfield  {author} {\bibinfo {author} {\bibfnamefont {T.}~\bibnamefont
  {Sch\"{a}fer}},\ }\bibfield  {title} {\bibinfo {title} {Generalized theory of
  diffusion based on kinetic theory},\ }\href@noop {} {\bibfield  {journal}
  {\bibinfo  {journal} {Phys. Rev. A}\ }\textbf {\bibinfo {volume} {94}},\
  \bibinfo {pages} {043644} (\bibinfo {year} {2016})}\BibitemShut {NoStop}%
\bibitem [{\citenamefont {Enss}(2013)}]{EnssTransvSpin}%
  \BibitemOpen
  \bibfield  {author} {\bibinfo {author} {\bibfnamefont {T.}~\bibnamefont
  {Enss}},\ }\bibfield  {title} {\bibinfo {title} {Transverse spin diffusion in
  strongly interacting {Fermi} gases},\ }\href@noop {} {\bibfield  {journal}
  {\bibinfo  {journal} {Phys. Rev. A}\ }\textbf {\bibinfo {volume} {88}},\
  \bibinfo {pages} {033630} (\bibinfo {year} {2013})}\BibitemShut {NoStop}%
\bibitem [{\citenamefont {Bardon}\ \emph {et~al.}(2014)\citenamefont {Bardon},
  \citenamefont {Beattie}, \citenamefont {Luciuk}, \citenamefont {Cairncross},
  \citenamefont {Fine}, \citenamefont {Cheng}, \citenamefont {Edge},
  \citenamefont {Taylor}, \citenamefont {Zhang}, \citenamefont {Trotzky},\ and\
  \citenamefont {Thywissen}}]{ThywissenTransvSpin}%
  \BibitemOpen
  \bibfield  {author} {\bibinfo {author} {\bibfnamefont {A.~B.}\ \bibnamefont
  {Bardon}}, \bibinfo {author} {\bibfnamefont {S.}~\bibnamefont {Beattie}},
  \bibinfo {author} {\bibfnamefont {C.}~\bibnamefont {Luciuk}}, \bibinfo
  {author} {\bibfnamefont {W.}~\bibnamefont {Cairncross}}, \bibinfo {author}
  {\bibfnamefont {D.}~\bibnamefont {Fine}}, \bibinfo {author} {\bibfnamefont
  {N.~S.}\ \bibnamefont {Cheng}}, \bibinfo {author} {\bibfnamefont {G.~J.~A.}\
  \bibnamefont {Edge}}, \bibinfo {author} {\bibfnamefont {E.}~\bibnamefont
  {Taylor}}, \bibinfo {author} {\bibfnamefont {S.}~\bibnamefont {Zhang}},
  \bibinfo {author} {\bibfnamefont {S.}~\bibnamefont {Trotzky}},\ and\ \bibinfo
  {author} {\bibfnamefont {J.~H.}\ \bibnamefont {Thywissen}},\ }\bibfield
  {title} {\bibinfo {title} {Transverse demagnetization dynamics of a unitary
  {Fermi} gas},\ }\href@noop {} {\bibfield  {journal} {\bibinfo  {journal}
  {Science}\ }\textbf {\bibinfo {volume} {344}},\ \bibinfo {pages} {722}
  (\bibinfo {year} {2014})}\BibitemShut {NoStop}%
\bibitem [{\citenamefont {{Luttinger}}\ and\ \citenamefont
  {{Ward}}(1960)}]{Luttinger1960}%
  \BibitemOpen
  \bibfield  {author} {\bibinfo {author} {\bibfnamefont {J.~M.}\ \bibnamefont
  {{Luttinger}}}\ and\ \bibinfo {author} {\bibfnamefont {J.~C.}\ \bibnamefont
  {{Ward}}},\ }\bibfield  {title} {\bibinfo {title} {Ground-state energy of a
  many-fermion system. ii},\ }\href@noop {} {\bibfield  {journal} {\bibinfo
  {journal} {Phys. Rev.}\ }\textbf {\bibinfo {volume} {118}},\ \bibinfo {pages}
  {1417} (\bibinfo {year} {1960})}\BibitemShut {NoStop}%
\bibitem [{\citenamefont {{Haussmann}}\ \emph {et~al.}(2007)\citenamefont
  {{Haussmann}}, \citenamefont {{Rantner}}, \citenamefont {{Cerrito}},\ and\
  \citenamefont {{Zwerger}}}]{Haussmann2007}%
  \BibitemOpen
  \bibfield  {author} {\bibinfo {author} {\bibfnamefont {R.}~\bibnamefont
  {{Haussmann}}}, \bibinfo {author} {\bibfnamefont {W.}~\bibnamefont
  {{Rantner}}}, \bibinfo {author} {\bibfnamefont {S.}~\bibnamefont
  {{Cerrito}}},\ and\ \bibinfo {author} {\bibfnamefont {W.}~\bibnamefont
  {{Zwerger}}},\ }\bibfield  {title} {\bibinfo {title} {Thermodynamics of the
  {BCS-BEC} crossover},\ }\href@noop {} {\bibfield  {journal} {\bibinfo
  {journal} {Phys. Rev. A}\ }\textbf {\bibinfo {volume} {75}},\ \bibinfo
  {pages} {023610} (\bibinfo {year} {2007})}\BibitemShut {NoStop}%
\bibitem [{\citenamefont {{Frank}}(2019)}]{Frank2019}%
  \BibitemOpen
  \bibfield  {author} {\bibinfo {author} {\bibfnamefont {B.}~\bibnamefont
  {{Frank}}},\ }\emph {\bibinfo {title} {Thermodynamics and Transport in Fermi
  Gases near Unitarity}},\ \href@noop {} {Ph.D. thesis},\ \bibinfo  {school}
  {Technische Universit\"{a}t M\"{u}nchen} (\bibinfo {year} {2019}),\ \bibinfo
  {note} {english}\BibitemShut {NoStop}%
\bibitem [{\citenamefont {{Bruun}}\ and\ \citenamefont
  {{Smith}}(2007)}]{BruunViscousNormalDamping}%
  \BibitemOpen
  \bibfield  {author} {\bibinfo {author} {\bibfnamefont {G.~M.}\ \bibnamefont
  {{Bruun}}}\ and\ \bibinfo {author} {\bibfnamefont {H.}~\bibnamefont
  {{Smith}}},\ }\bibfield  {title} {\bibinfo {title} {Shear viscosity and
  damping for a {Fermi} gas in the unitary limit},\ }\href@noop {} {\bibfield
  {journal} {\bibinfo  {journal} {Phys. Rev. A}\ }\textbf {\bibinfo {volume}
  {75}},\ \bibinfo {pages} {043612} (\bibinfo {year} {2007})}\BibitemShut
  {NoStop}%
\bibitem [{\citenamefont {{Bluhm}}\ and\ \citenamefont
  {{Sch\"afer}}(2016)}]{BluhmSchaeferModIndep}%
  \BibitemOpen
  \bibfield  {author} {\bibinfo {author} {\bibfnamefont {M.}~\bibnamefont
  {{Bluhm}}}\ and\ \bibinfo {author} {\bibfnamefont {T.}~\bibnamefont
  {{Sch\"afer}}},\ }\bibfield  {title} {\bibinfo {title} {Model-independent
  determination of the shear viscosity of a trapped unitary {Fermi} gas:
  Application to high-temperature data},\ }\href@noop {} {\bibfield  {journal}
  {\bibinfo  {journal} {Phys. Rev. A}\ }\textbf {\bibinfo {volume} {116}},\
  \bibinfo {pages} {115301} (\bibinfo {year} {2016})}\BibitemShut {NoStop}%
\bibitem [{\citenamefont {{Bluhm}}\ \emph {et~al.}(2017)\citenamefont
  {{Bluhm}}, \citenamefont {{Hou}},\ and\ \citenamefont
  {{Sch\"afer}}}]{BluhmSchaeferLocalViscosity}%
  \BibitemOpen
  \bibfield  {author} {\bibinfo {author} {\bibfnamefont {M.}~\bibnamefont
  {{Bluhm}}}, \bibinfo {author} {\bibfnamefont {J.}~\bibnamefont {{Hou}}},\
  and\ \bibinfo {author} {\bibfnamefont {T.}~\bibnamefont {{Sch\"afer}}},\
  }\bibfield  {title} {\bibinfo {title} {Determination of the density and
  temperature dependence of the shear viscosity of a unitary {Fermi} gas based
  on hydrodynamic flow},\ }\href@noop {} {\bibfield  {journal} {\bibinfo
  {journal} {Phys. Rev. Lett.}\ }\textbf {\bibinfo {volume} {119}},\ \bibinfo
  {pages} {065302} (\bibinfo {year} {2017})}\BibitemShut {NoStop}%
\bibitem [{\citenamefont {{Braby}}\ \emph {et~al.}(2010)\citenamefont
  {{Braby}}, \citenamefont {{Chao}},\ and\ \citenamefont
  {{Sch\"afer}}}]{BrabySchaeferThermalCond}%
  \BibitemOpen
  \bibfield  {author} {\bibinfo {author} {\bibfnamefont {M.}~\bibnamefont
  {{Braby}}}, \bibinfo {author} {\bibfnamefont {J.}~\bibnamefont {{Chao}}},\
  and\ \bibinfo {author} {\bibfnamefont {T.}~\bibnamefont {{Sch\"afer}}},\
  }\bibfield  {title} {\bibinfo {title} {Thermal conductivity and sound
  attenuation in dilute atomic {Fermi} gases},\ }\href@noop {} {\bibfield
  {journal} {\bibinfo  {journal} {Phys. Rev. A}\ }\textbf {\bibinfo {volume}
  {82}},\ \bibinfo {pages} {033619} (\bibinfo {year} {2010})}\BibitemShut
  {NoStop}%
\bibitem [{\citenamefont {{Li}}\ \emph {et~al.}(2024)\citenamefont {{Li}},
  \citenamefont {{Huang}},\ and\ \citenamefont {{Thomas}}}]{XiangDensityshift}%
  \BibitemOpen
  \bibfield  {author} {\bibinfo {author} {\bibfnamefont {X.}~\bibnamefont
  {{Li}}}, \bibinfo {author} {\bibfnamefont {J.}~\bibnamefont {{Huang}}},\ and\
  \bibinfo {author} {\bibfnamefont {J.~E.}\ \bibnamefont {{Thomas}}},\
  }\bibfield  {title} {\bibinfo {title} {Universal density shift coefficients
  for the thermal conductivity and shear viscosity of a unitary {Fermi} gas},\
  }\href@noop {} {\bibfield  {journal} {\bibinfo  {journal} {Phys. Rev. Res.}\
  }\textbf {\bibinfo {volume} {6}},\ \bibinfo {pages} {L042021} (\bibinfo
  {year} {2024})}\BibitemShut {NoStop}%
\bibitem [{\citenamefont {Patel}\ \emph {et~al.}(2020)\citenamefont {Patel},
  \citenamefont {Yan}, \citenamefont {Mukherjee}, \citenamefont {Fletcher},
  \citenamefont {Struck},\ and\ \citenamefont {Zwierlein}}]{MZSound}%
  \BibitemOpen
  \bibfield  {author} {\bibinfo {author} {\bibfnamefont {P.~B.}\ \bibnamefont
  {Patel}}, \bibinfo {author} {\bibfnamefont {Z.}~\bibnamefont {Yan}}, \bibinfo
  {author} {\bibfnamefont {B.}~\bibnamefont {Mukherjee}}, \bibinfo {author}
  {\bibfnamefont {R.~J.}\ \bibnamefont {Fletcher}}, \bibinfo {author}
  {\bibfnamefont {J.}~\bibnamefont {Struck}},\ and\ \bibinfo {author}
  {\bibfnamefont {M.~W.}\ \bibnamefont {Zwierlein}},\ }\bibfield  {title}
  {\bibinfo {title} {Universal sound diffusion in a strongly interacting
  {Fermi} gas},\ }\href@noop {} {\bibfield  {journal} {\bibinfo  {journal}
  {Science}\ }\textbf {\bibinfo {volume} {370}},\ \bibinfo {pages} {1222}
  (\bibinfo {year} {2020})}\BibitemShut {NoStop}%
\bibitem [{\citenamefont {Frank}\ \emph {et~al.}(2020)\citenamefont {Frank},
  \citenamefont {Zwerger},\ and\ \citenamefont {Enss}}]{EnssTransport}%
  \BibitemOpen
  \bibfield  {author} {\bibinfo {author} {\bibfnamefont {B.}~\bibnamefont
  {Frank}}, \bibinfo {author} {\bibfnamefont {W.}~\bibnamefont {Zwerger}},\
  and\ \bibinfo {author} {\bibfnamefont {T.}~\bibnamefont {Enss}},\ }\bibfield
  {title} {\bibinfo {title} {Quantum critical thermal transport in the unitary
  {Fermi} gas},\ }\href@noop {} {\bibfield  {journal} {\bibinfo  {journal}
  {Phys. Rev. Research}\ }\textbf {\bibinfo {volume} {2}},\ \bibinfo {pages}
  {023301} (\bibinfo {year} {2020})}\BibitemShut {NoStop}%
\end{thebibliography}
\end{document}